\documentclass[twocolumn,english,aps,prb,amsmath,amssymb,nofootinbib,superscriptaddress,floatfix,preprintnumbers]{revtex4}
\usepackage[T1]{fontenc}
\usepackage[latin9]{inputenc}
\usepackage{amsmath}
\usepackage{graphicx}
\usepackage{amssymb}

\makeatletter
\@ifundefined{textcolor}{}
{%
 \definecolor{BLACK}{gray}{0}
 \definecolor{WHITE}{gray}{1}
 \definecolor{RED}{rgb}{1,0,0}
 \definecolor{GREEN}{rgb}{0,1,0}
 \definecolor{BLUE}{rgb}{0,0,1}
 \definecolor{CYAN}{cmyk}{1,0,0,0}
 \definecolor{MAGENTA}{cmyk}{0,1,0,0}
 \definecolor{YELLOW}{cmyk}{0,0,1,0}
 }

\@ifundefined{definecolor}
 {\usepackage{color}}{}

\makeatother

\usepackage{babel}

\begin{document}

\title{Decay rates for topological memories encoded with Majorana fermions}

\author{G. Goldstein}

\affiliation{Physics Department, Harvard University, Cambridge MA 02138, USA}

\author{C. Chamon}

\affiliation{Physics Department, Boston University, Boston MA 02215 USA}
\begin{abstract}
Recently there have been numerous proposals to create Majorana zero
modes in solid state heterojunctions, superconducting wires and optical
lattices. Putatively the information stored in qubits constructed
from these modes is protected from various forms of decoherence. Here
we present a generic method to study the effect of external perturbations
on these modes. We focus on the case where there are no interactions
between different Majorana modes either directly or through intermediary
fermions. To quantify the rate of loss of the information stored in
the Majorana modes we study the two-time correlators for qubits built
from them. We analyze a generic gapped fermionic environment (bath)
interacting via tunneling with different components of the qubit (different
Majorana modes). We present examples with both static and dynamic
perturbations (noise), and using our formalism we derive a rate of
information loss, for Majorana memories, that depends on the spectral
density of both the noise and the fermionic bath.


\end{abstract}
\maketitle

\section{Introduction\label{sec:Introduction}}

Topological quantum computation requires the existence of topologically
ordered states whose low energy excitations follow non-Abelian statistics.
The subspace of states corresponding to a fixed number of quasiparticles
is degenerate, to an exponential precision, in the separation between
quasiparticles, and an exchange of the positions of these anyonic
excitations, also known as braiding, leads to a unitary transformation
within this low energy subspace. These unitary operations are insensitive
to the exact path used to perform the braiding operation and in many
cases, for an appropriate encoding, braiding operations correspond
to {}``standard'' one- and two-qubit gates within the low energy
subspace. These operations can be used as building blocks for fault
tolerant quantum computation.

There are many candidate systems for experimental realizations of
topological phases of matter with these properties. There is preliminary
evidence that the $\nu=5/2$ fractional quantum Hall state may have
non-Abelian excitations \cite{key-23,key-24,key-25}. Spin-triplet
$p_{x}+ip_{y}$ pairing superfluidity occurs in the A-phase of $\mathrm{^{3}He}$
\cite{key-26,key-27} and in strontium ruthenates \cite{key-28,key-29,key-30,key-31,key-32,key-83},
in which half quantum vortices would be non-Abelian \cite{key-33,key-3}.
There are also proposals to realize chiral p-wave superconductors
in ultra-cold atom systems \cite{key-21,key-35,key-34}. Furthermore
there have been many advances towards producing topological states
of matter in layered heterojunction systems \cite{key-36,key-37,key-38,key-15,key-2,key-4,key-5,key-7,key-8,key-16,key-17,key-18,key-20,key-6}.

Virtually all current experimentally viable proposals for platforms
for topological quantum computation only support Ising type anyons
which are carried by Majorana fermion modes. Colloquially speaking
these fermions are half of a regular fermion. More precisely they
are self-adjoint operators $\gamma_{i}$ which can be written as a
sum of an annihilation and creation operator for one fermion mode
and which satisfy the algebra: \begin{equation}
\left\{ \gamma_{i},\,\gamma_{j}\right\} =2\delta_{ij},\,\gamma_{i}^{\dagger}=\gamma_{i}\label{eq:BasicDefintion}\end{equation}
 Any two Majorana fermion operators can be combined into a regular
fermion mode $c$ and its adjoint $c^{\dagger}$ via $c=\frac{1}{2}\left(\gamma_{1}+i\gamma_{2}\right)$
and $c^{\dagger}=\frac{1}{2}\left(\gamma_{1}-i\gamma_{2}\right)$.

The topological qubit is made up of four spin polarized MBSs $\gamma_{1},\,\gamma_{2},\,\gamma_{3}$
and $\gamma_{4}$ \cite{key-43}. These can be combined into two sets
of creation and annihilation operators:\begin{eqnarray}
\begin{array}{rcl}
c_{1} & = & \frac{1}{2}\left(\gamma_{1}+i\gamma_{2}\right)\end{array} &  & \begin{array}{rcl}
c_{1}^{\dagger} & = & \frac{1}{2}\left(\gamma_{1}-i\gamma_{2}\right)\end{array}\label{eq:Operators}\\
\begin{array}{rcl}
c_{2} & = & \frac{1}{2}\left(\gamma_{3}+i\gamma_{4}\right)\end{array} &  & \begin{array}{rcl}
c_{2}^{\dagger} & = & \frac{1}{2}\left(\gamma_{3}-i\gamma_{4}\right)\end{array}\nonumber \end{eqnarray}
For the logical basis it is convenient to work in the even fermion
parity subspace. The qubit basis can be chosen to be $\left|+_{L}\right\rangle \equiv\left|00\right\rangle $
and $\left|-_{L}\right\rangle \equiv\left|11\right\rangle $ where
the $0$'s and $1$'s refer to the occupation numbers relative to
the complex fermion operators in Eq. (\ref{eq:Operators}). Because
of fermion parity conservation, any operation that does not entangle
the states with the environment cannot mix even and odd fermion parity
states for the qubits. As such, all gates acting on the topological
qubit should not take the system out of the logical subspace. Furthermore
all the operators of the single spin Clifford group may be produced
by braiding the four vortices of our qubit leading to potentially
topologically protected gates \cite{key-14}. In particular the various
single qubit operations in our logic basis may be conveniently written
in terms of the Majorana operators. For future use we note that in
this encoding \begin{equation}
\sigma^{z}=-i\gamma_{1}\gamma_{2},\;\sigma^{x}=-i\gamma_{2}\gamma_{3},\;\sigma^{y}=i\gamma_{1}\gamma_{3}.\label{eq:OperatorsSigma}\end{equation}
 Here all the sigma matrices are with respect to the logic basis $\left|+_{L}\right\rangle $
and $\left|-_{L}\right\rangle $. We will primarily be interested
in correlators of the form $\left\langle \sigma^{\mathrm{z}}\left(0\right)\sigma^{\mathrm{z}}\left(\mathrm{T}\right)\right\rangle =-\left\langle \gamma_{1}\left(0\right)\gamma_{2}\left(0\right)\gamma_{1}\left(\mathrm{T}\right)\gamma_{2}\left(\mathrm{T}\right)\right\rangle $.
We will proceed to calculate these below.


The Majorana operators are zero modes of some mean field Hamiltonian
$\left[{H}_{\mathrm{MF}},\,\gamma_{i}\right]=0$ so it can be argued
that these modes are protected from decoherence as the mean field
Hamiltonian when restricted to the subspace generated by these modes
is zero. One of the open tasks of topological quantum computation
is associated with understanding the extent of this protection. This
is the subject of this paper.

\section{Summary of main ideas}

In this section we outline the setup of the rest of the paper. We
present the relevant Hamiltonian and discuss its basic properties.
We describe the type of qubit we will focus on in the text, a localized
Majorana mode, and give an overview of some other encodings we shall
not consider in this paper. We describe the kinds of calculations
of memory coherence we are going to do in this paper. We also give
a Section by Section outline.

We begin our discussion with relevant Hamiltonians. The Majorana fermions
interact with the external environment via tunneling type Hamiltonians.
On symmetry grounds, for a single Majorana mode, any such interaction
may be written as: \begin{equation}
{H_{\mathrm{int}}}=\gamma\int{d^{d}r\left[u_{0}(\vec{r})\,\Phi^{\dagger}\left(\vec{r}\right)\Psi^{\dagger}\left(\vec{r}\right)-u_{0}^{*}(\vec{r})\,\Psi\left(\vec{r}\right)\Phi\left(\vec{r}\right)\right]}.\label{eq:TunnelingHamiltonian}\end{equation}
 Here $u_{0}(\vec{r})$ is the localized mode function associated
with the Majorana bound state, $\Phi(\vec{r})$ is any local bosonic
field, which in the simplest case is a tunneling amplitude (complex
number) and $\Psi(\vec{r})$ is a regular (complex) fermion field.
In this paper we will analyze multiple Majorana fermions coupled to
different types of environments via Hamiltonians of the form given
in Eq. (\ref{eq:TunnelingHamiltonian}). Furthermore the fermions
in the bath will always be assumed to be gapped, for example, electrons
in an insulating or superconducting material (environments composed
of gapless fermions, instead, would obviously lead to decoherence).

There are many examples of microscopic situations where Hamiltonians
of the form given in Eq.~(\ref{eq:TunnelingHamiltonian}) arise,
one is as follows. If one writes the mode expansion of the electron
creation and annihilation operators in the (superconducting) system
of interest, one finds that: \begin{eqnarray}
\left(\begin{array}{l}
\psi\left(\vec{r},\, t\right)\\
\psi^{\dagger}\left(\vec{r},\, t\right)\end{array}\right) & = & \gamma\left(\begin{array}{l}
u_{0}\left(\vec{r}\right)\\
u_{0}^{*}\left(\vec{r}\right)\end{array}\right)\nonumber \\
 & + & \sum_{|E|>0}a_{E}\; e^{-iEt}\,\left(\begin{array}{l}
u_{E}\left(\vec{r}\right)\\
v_{E}\left(\vec{r}\right)\end{array}\right)\;.\label{eq:Decopositionfield-intro}\end{eqnarray}
 Here $a_{E}$ stands for the eigenoperators of the BdG equations,
with non-zero energies, while $u_{E}$ and $v_{E}$ are the components
of the corresponding eigenmode of the BdG equations. $\gamma$ is
the Majorana fermion corresponding to the zero energy mode. Now consider
an insulating substrate below a system which may be described by Eq.
(\ref{eq:Decopositionfield-intro}) above. A concrete example is given
by the bulk of a topological insulator in tunneling contact with a
superconductor as shown in Ref. ~\onlinecite{key-11}. For a static
Hamiltonian the bulk and surface states are orthogonalized, but dynamical
effects such as phonons or two-level defect systems can alter the
original Hamiltonian and turn on a hybridization. This perturbation
takes the form of a tunneling between the electrons: ${H_{\mathrm{int}}}=\int d^{d}r\;\Phi\left(\vec{r}\right)\;\Psi^{\dagger}\left(\vec{r}\right)\,\Psi\left(\vec{r}\right)+{\rm h.\, c.}$,
where $\Phi\left(\vec{r}\right)$ controls the amplitude of fluctuations
of the tunneling coupling. $\Phi\left(\vec{r}\right)$ can be due
to phonons, two-level systems, or even classical sources of noise.
The electrons $\Psi\left(\vec{r}\right)$ come from the insulating
(gapped) system, which comprise the fermionic component of our bath.
This illustrates one of the many ways to arrive at Hamiltonians of
the form Eq.~(\ref{eq:TunnelingHamiltonian}).

The coupling Hamiltonian that is derived in the paragraph above is
local. The terms in Eq.~(\ref{eq:TunnelingHamiltonian}) are local
and couple to only one Majorana mode, with no long distance coupling
between the modes of any form. In this paper we shall focus on sets
of baths that couple to each Majorana individually. We would like
to stress now and henceforth that even by coupling to individual modes,
one at a time (with no cross mode coupling), the bath can be very
damaging, in many cases leading to zero coherence for long times.

Below, we look at decoherence by analyzing qubit correlations such
as $\left\langle \sigma^{\mathrm{z}}\left(0\right)\sigma^{\mathrm{z}}\left(\mathrm{T}\right)\right\rangle =-\left\langle \gamma_{1}\left(0\right)\gamma_{2}\left(0\right)\gamma_{1}\left(\mathrm{T}\right)\gamma_{2}\left(\mathrm{T}\right)\right\rangle $,
which, as we show in this paper, factorizes when the baths that couple
to each Majorana are uncorrelated with one another: \begin{eqnarray}
\left\langle \sigma^{z}\left(0\right)\sigma^{z}\left(\mathrm{T}\right)\right\rangle  & = & \left\langle \gamma_{1}\left(0\right)\gamma_{1}\left(\mathrm{T}\right)\right\rangle \times\left\langle \gamma_{2}\left(0\right)\gamma_{2}\left(\mathrm{T}\right)\right\rangle .\end{eqnarray}
 Thus, even though the qubit is defined non-locally using spatially
separated Majorana fermions, below we will show that the decay of
the memory is controlled by the product of the two-time correlations
of the separate Majorana modes. It then suffices to understand the
effect of the bath on each Majorana fermion separately.

\begin{figure}[htbp]
\begin{centering}
\includegraphics[scale=1.2]{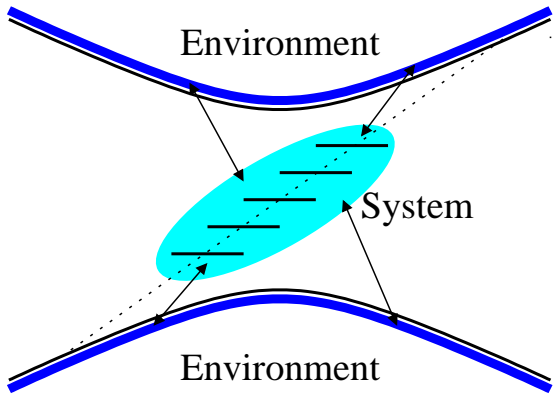} 
\par\end{centering}

\caption{\label{fig-SystemEnvironmenttunneling} Depiction of the separation
between system and bath degrees of freedom. For infinite baths, one
cannot construct a local operator of the form Eq.~(\ref{eq:parity-string}),
one that is a product of a finite number of terms. If the product
is truncated, parity leaks into the bath.}

\end{figure}

At this point its worthwhile to stress that the qubit encoding given
above is not unique. A particularly interesting example of a different
encoding, given by Akhmerov~\cite{key-1}, is a fermion parity protected
encoding. There, the qubit is made from fermion parity preserving
operators: \begin{equation}
\tilde{\gamma}=\gamma\;\prod_{i}(1-2\, c_{i}^{\dagger}\, c_{i}^{})\label{eq:parity-string}\end{equation}
 that commute with both the tunneling Hamiltonian and the Hamiltonian
for the environment. Here the $c_{i}^{}$ are the operators in the
mode expansion of the fermionic $\Psi\left(\vec{r}\right)$ field
in the bath ($i$ here labels the mode, which can be momentum, for
example). For a finite system, such as mid gap Carroli Matricon deGennes
states in vortex cores, this compound qubit is very efficient. However
we stress that, in the presence of a bath (say made by continuum states),
the construction of an operator that is protected because of parity
conservation requires a product of infinitely many operators: which
is not practical or easily experimentally measurable. One could also
truncate the product so as to account for a system, and the terms
omitted are those assigned to the bath, as depicted in Fig.~\ref{fig-SystemEnvironmenttunneling}.
In this case, however, because the operator lacks degrees of freedom
assigned to the bath, parity can leak to the environment decohering
the qubit. As such we will ignore all {}``compound'' encodings for
the rest of the paper. %

Finally, we would also like to mention that the above scheme, with
simple, non-compound, Majorana encoding, generalizes to multiple qubits.
One possible encoding (though not the most economical) is to use four
vortices and as such four Majorana modes per qubit. For this and any
other encoding all possible correlators for the quantum memory may
be expressed as expectation values of various products of Majorana
operators\cite{key-89}. All quantum coherences for our qubits may
then be computed by studying Majorana mode correlators which we study
below.

In carrying out this program, we will analyze two distinct types of
environments: the first is when couplings $\Phi(\vec{r})$ change
suddenly but remain static thereafter, and the second when the environment
changes dynamically. We show that that in the static environment case
the tunneling Hamiltonian merely leads to a finite depletion of the
Majorana two-time correlations. In this case, much of the information
stored in these modes survives for arbitrarily long times. 

More generally, for dynamic environments, we obtain an expression
for the rate of loss of information stored in the Majorana operators
that depends on the spectral density of the noise and of the fermionic
bath. We present several examples of noise that can be studied essentially
exactly, for instance classical telegraphic noise, as well as both
classical and quantum Gaussian fluctuations.

The results in the paper are presented as follows: 
\begin{itemize}
\item In Section \ref{sec:Dynamics} we present general considerations involving
the coherence properties of Majorana modes. We show that under reasonably
generic initial conditions the coherence of the Majorana modes does
not depend on their initial states. Furthermore we show that the two
time correlation functions, coherences, factorize as a product over
coherences for individual Majorana modes, that make up the quantum
memory, interacting with their individual environments. As such we
may reduce the problem of the coherence of the quantum memory to the
problem of the coherence of one Majorana mode in tunneling contact
with a (gapped) fermionic reservoir. 
\item In Section \ref{sec:Keldysh} we take a first step towards a calculation
of the coherence of a single Majorana mode. We begin by describing
the Keldysh technique relevant to Majorana modes. We present combinatorial
tricks that make is possible to efficiently convert Keldysh computations
using a mixture of Majorana and regular fermionic modes into a more
familiar computation which uses only regular fermion modes. We then
present an example where, for simplicity, we treat the fermions in
the bath as free (non-relaxing approximation). We also present a general
formula for the coherence of a Majorana qubit that is used several
times in the remaining analysis. 
\item In Section \ref{sec:Fluctuating-Hamiltonians} we present several
related classical models for the fluctuations of the bath. We solve
these models essentially exactly, by mapping the problem of the coherence
of a single Majorana mode to the problem of a particle undergoing
classical diffusion. We use this technique to study classical fluctuations
of the tunneling amplitudes and energy levels of the reservoir (we
primarily focus on Gaussian fluctuations). In all cases we find decoherence
with a rate that depends on the spectral density of the fluctuations
in the reservoir. In many cases the decoherence due to an individual
fermion mode has a power law time dependence but it will turn out
that a bath made of many weakly interacting modes leads to exponential
decay of coherence for intermediate times. 

\item In Section \ref{sec:Conclusions} we conclude. In light of the results
we obtain in this paper, we critically examine the degree in which
quantum memories can be encoded using Majorana fermions when these
are in contact with a dynamical environment. We show that the coherence
of the Majorana mode is controlled by the coherence of the bath it
interacts with. 
\item In Appendix \ref{sec:QuadraticHamiltonian} we compute exact dressed
zero modes for static quadratic Hamiltonians, which we use to verify
the validity of our results in Section~\ref{sec:Keldysh}. In Appendix
\ref{sub:Quantum-Fluctuations} we present a rather technical calculation
of a Majorana mode interacting with a fermionic bath with fully quantum
mechanical Gaussian fluctuations. To leading order we find a decay
similar to classical computations. In Appendix \ref{sec:Tedious-Calculations}
we present various technical calculations, used throughout the rest
of the text. In particular, in Appendix \ref{sub:Parity-Eigenvalues-(Coding}
we show that our results are independent of coding subspace, in Appendix
\ref{sec:Justification} we present some technical arguments (which
are used in Section \ref{sec:Fluctuating-Hamiltonians}) in favor
of weak (negligible) coupling of the fluctuation for the various fermionic
modes. In the rest of the appendix we derive formulas used in the
main text. 
\end{itemize}

\section{Dynamics \label{sec:Dynamics}}

We begin with a study of the general properties of the dynamics of
a system of Majorana modes. We will focus on a computation of correlators
involving Majorana operators. This will allow us to study the coherence
properties of a topological quantum memory which is based on qubits
made up of localized zero energy modes. In this Section we will adhere
to very general Hamiltonians and we will study only properties that
are essentially independent of the form of this Hamiltonian. This
will set us up for studies of specific types of Hamiltonians in Section
\ref{sec:Keldysh}. From the outset, we would like to specify the
initial conditions or equivalently the density matrix when the system
is initialized at $t=0$. We will assume that initially the density
matrix factorizes into a product of the form:\begin{equation}
\rho_{\mathrm{tot}}=\rho_{\mathrm{Maj}}\otimes\prod_{i}\rho_{\mathrm{env_{i}}}\label{eq:DensityMatrix}\end{equation}
 Here $\rho_{\mathrm{tot}}$ is the density matrix for the entire
system, while $\rho_{\mathrm{Maj}}$ represents and arbitrary non-equilibrium
density matrix for the Majorana modes. The $\rho_{\mathrm{env_{i}}}$
are arbitrary, not-necessarily equilibrium, density matrices for the
environments of the individual Majorana modes. No specific {}``ensemble''
is assumed. This form is a reasonable, consistent assumption for the
initial states of system plus bath, particularly so, as many experimental
methods of initialization produce such states.

For our qubit memory persistence between times $t_{1}$ and $t_{2}$
is captured by the two-time correlators such as $\left\langle \sigma^{z}\left(t_{1}\right)\sigma^{z}\left(t_{2}\right)\right\rangle $.
We note that, because the initial, $t=0$, state breaks time-translation
invariance, generically these correlators are functions of both $t_{1}$
and $t_{2}$. Here we shall focus specifically on correlations, like
$\left\langle \sigma^{z}\left(0\right)\sigma^{z}\left(\mathrm{T}\right)\right\rangle $,
between the state prepared at $t=0$ and the state at a later time
time $t=\mathrm{T}$ which characterize the degree to which the information
encoded in the qubit at the initial time survives interaction with
the bath when it is retrieved at a later time $\mathrm{T}$.

The key results of this section, which are used repeatedly later in
the text, may be summarized by saying that even though the factorization
form given in Eq.~(\ref{eq:DensityMatrix}) does not survive Hamiltonian
evolution the expectation values of various correlators like $\left\langle \sigma^{z}\left(0\right)\sigma^{z}\left(\mathrm{T}\right)\right\rangle $
or equivalently products of Majorana fermions, to be defined precisely
in Eqs. (\ref{eq:SimplifedCohrence}) and (\ref{eq:FinalCorrelatorForm})
below, do factorize into products of expectation values for individual
Majorana modes. This factorization survives for arbitrary times.

\subsection{\label{sub:General-Ideas}General ideas}

We will consider a set of Majorana modes each interacting with its
own fermionic environment, see Eq. (\ref{eq:DensityMatrix}). We will
see that there is decoherence even without direct interactions between
different Majorana modes or between their respective environments.
One can show that, in the limit when the spatial separation between
the Majorana modes is large, the case when multiple Majorana modes
interact with a common fermionic bath reduces to the case of uncorrelated
non-interacting baths (see Appendix \ref{sub:Cross-Correlations-between}).
The Hamiltonian pertinent to each mode may be written as:\begin{eqnarray}
H_{\alpha} & = & \sum_{i=1}^{N_{\alpha}}\gamma_{\alpha}\left[\left(B_{i,\alpha}^{\phantom{\dagger}}\, c_{i,\alpha}^{\phantom{\dagger}}-c_{i,\alpha}^{\dagger}\, B_{i,\alpha}^{\dagger}\right)\right.\nonumber \\
 & + & \left.H_{\alpha}^{\mathrm{bath}}\left(\{c_{i,\alpha}^{\phantom{\dagger}},c_{i,\alpha}^{\dagger},B_{i,\alpha}^{\phantom{\dagger}},B_{i,\alpha}^{\dagger}\}\right)\right]\;.\label{eq:GenericHamiltonian}\end{eqnarray}

Here $B_{i,\alpha}$ are some bosonic modes and $\alpha=\left\{ 1,2,...\right\} $
labels the Majorana modes. The total Hamiltonian is given by $H=\sum_{\alpha}H_{\alpha}$.
We will be interested in correlators of the form $\left\langle \gamma_{\alpha_{1}}\;\gamma_{\alpha_{2}}\dots\gamma_{\alpha_{k}}\;\gamma_{\alpha_{1}}\left(t_{1}\right)\;\gamma_{\alpha_{2}}\left(t_{2}\right)\dots\gamma_{\alpha_{\mathrm{k}}}\left(t_{k}\right)\right\rangle $.
Here all operators are in the Heisenberg picture, and $\gamma_{\alpha}\left(t\right)$
is given by \begin{equation}
\gamma_{\alpha}(t)=\left(\widetilde{\mathcal{T}}\; e^{i\int_{0}^{t}H_{\alpha}\left(\tau\right)d\tau}\right)\;\;\gamma_{\alpha}\;\;\left(\mathcal{T}\; e^{-i\int_{0}^{t}H_{\alpha}\left(\tau\right)d\tau}\right)\;,\label{eq:gamma-t}\end{equation}
 where ${\mathcal{T}}$ and $\widetilde{\mathcal{T}}$ stand for time-ordered
and anti-time-ordered products, respectively. Notice that $\gamma_{\alpha}\left(t\right)=\gamma_{\alpha}^{\dagger}\left(t\right)$
at all times.

Now, by Taylor expanding the time-ordered and anti-time-ordered exponentials
in Eq.~(\ref{eq:gamma-t}), taking various commutators, grouping
terms and using the fact that $\gamma_{\alpha}^{2}=1$, we may write
that \begin{equation}
\gamma_{\alpha}\left(t\right)=\gamma_{\alpha}\,\mathcal{B}_{\alpha}\left(t\right)+\mathcal{F}_{\alpha}\left(t\right),\label{eq:Key Form}\end{equation}
 with $\mathcal{B}_{\alpha}\left(t\right)$ and $\mathcal{F}_{\alpha}\left(t\right)$
having no factors of $\gamma_{\alpha}$. Because $\gamma_{\alpha}\left(t\right)$
must be fermionic (this can be seen from the fact that the Hamiltonian
and all its powers are bosonic) we may deduce that $\mathcal{B}_{\alpha}\left(t\right)$
and $\mathcal{F}_{\alpha}\left(t\right)$ are, respectively, bosonic
and fermionic operators. By the conservation of fermion parity we
know that the expectation value of any operator $\left\langle \mathcal{F}_{\alpha}\left(t\right)\right\rangle =0$.
Finally, because $\gamma_{\alpha}\left(t\right)$ is Hermitian, it
also follows from the properties above that $\mathcal{B}_{\alpha}\left(t\right)$
and $\mathcal{F}_{\alpha}\left(t\right)$ are Hermitian as well.

Now, it follows that \begin{eqnarray}
\left\langle \gamma_{\alpha}\;\gamma_{\alpha}\left(t\right)\right\rangle  & = & \left\langle \mathcal{B}_{\alpha}\left(t\right)\right\rangle +\left\langle \gamma_{\alpha}\,\mathcal{F}_{\alpha}\left(t\right)\right\rangle \nonumber \\
 & = & \left\langle \mathcal{B}_{\alpha}\left(t\right)\right\rangle +\left\langle \gamma_{\alpha}\right\rangle \,\left\langle \mathcal{F}_{\alpha}\left(t\right)\right\rangle \nonumber \\
 & = & \left\langle \mathcal{B}_{\alpha}\left(t\right)\right\rangle \;,\label{eq:SimplifedCohrence}\end{eqnarray}
 where we used going from the first to the second line of Eq.~(\ref{eq:SimplifedCohrence})
that the environments and the Majorana states are initially disentangled
so expectation values factorize. Note that this comes about because
in the Heisenberg picture the expectation values for operators are
taken with respect to the initial state, at $t=0$. For the third
line we have used that the expectation value of any fermionic operator
$\left\langle \mathcal{F}_{\alpha}\left(t\right)\right\rangle $ should
be zero. Note that because $\mathcal{B}_{\alpha}\left(t\right)$ is
Hermitian this implies that $\left\langle \gamma_{\alpha}\;\gamma_{\alpha}\left(t\right)\right\rangle \in\mathbb{R}$.

\begin{widetext} The following factorization formula can be similarly
showed: \begin{eqnarray}
\left\langle \gamma_{\alpha_{1}}\dots\gamma_{\alpha_{\mathrm{k}}}\gamma_{\alpha_{1}}\left(t_{1}\right)\dots\gamma_{\alpha_{\mathrm{k}}}\left(t_{k}\right)\right\rangle  & = & \left(-1\right)^{k\left(k-1\right)/2}\;\;\prod_{j=1}^{k}\left\langle \mathcal{B}_{\alpha_{\mathrm{j}}}\left(t_{j}\right)\right\rangle \nonumber \\
 & = & \left(-1\right)^{k\left(k-1\right)/2}\;\;\prod_{j=1}^{k}\left\langle \gamma_{\alpha_{j}}\;\gamma_{\alpha_{j}}\left(t_{j}\right)\right\rangle \;,\label{eq:FinalCorrelatorForm}\end{eqnarray}
 for distinct ${\alpha_{j}}$, $j=1,\dots,k$. To show this expression,
one uses Eq.~(\ref{eq:Key Form}) and again that the expectation
values are computed with respect to the initial density matrix given
in Eq.~(\ref{eq:DensityMatrix}) which has the property that the
environments are uncorrelated with each other and with the initial
Majorana states. We see that this factorization formula is independent
of the initial state of the density matrix of the bath. As such our
formalism captures highly non-equilibrium initial conditions.

\subsection{\label{sub:Qubit-memory-correlations}Qubit memory correlations}

The degree of persistence of memories assembled using Majorana fermions
can be quantified by the correlation between the qubit state, encoded
as in Eq.~(\ref{eq:OperatorsSigma}), at two times $0,\,\mathrm{T}$:
\begin{eqnarray}
\left\langle \sigma^{z}\left(0\right)\sigma^{z}\left(\mathrm{T}\right)\right\rangle  & = & -\left\langle \gamma_{1}\left(0\right)\gamma_{2}\left(0\right)\gamma_{1}\left(\mathrm{T}\right)\gamma_{2}\left(\mathrm{T}\right)\right\rangle \nonumber \\
 & = & \;\;\;\left\langle \gamma_{1}\left(0\right)\gamma_{1}\left(\mathrm{T}\right)\right\rangle \times\left\langle \gamma_{2}\left(0\right)\gamma_{2}\left(\mathrm{T}\right)\right\rangle .\label{eq:qubit-memory-corr}\end{eqnarray}

Notice that the factorization implies that, even though the qubit
is defined non-locally using two spatially separated Majorana fermions,
the decay of the memory is controlled by the product of the two-time
correlations of the two separate Majorana modes. In particular, the
decoherence rate is independent of the initial state of the quantum
memory (that is correlators of the form $\left\langle \gamma_{1}\gamma_{2}\right\rangle $
do not enter the result).

\vspace{0.3cm}
 \end{widetext}

Thus in the case of uncoupled well separated Majorana modes each interacting
with its own environment the task of determining the persistence of
topological quantum memories based on Majorana fermions is reduced
to the calculation of the coherences $\left\langle \gamma_{\alpha}\left(0\right)\gamma_{\alpha}\left(\mathrm{T}\right)\right\rangle $
in the presence of different fermionic environments. We carry out
this program henceforth.


\section{Keldysh calculation of coherence \label{sec:Keldysh}}

We now proceed to describe the technical details associated with studying
dynamics. For generality and later use we will study both static and
time dependent Hamiltonians. Based on the discussion given in Section
\ref{sec:Dynamics} for the purposes of computing coherences it will
be sufficient to focus on a single Majorana mode. As such we will
drop the subscript $\alpha$, see Eq. (\ref{eq:GenericHamiltonian}),
henceforth.

\subsection{\label{sub:General-Observations}General Observations}

We will convert the computation of the Majorana correlations into
a Keldysh calculation carried out using only the bosons and regular
complex fermions inside the reservoir. (For a review of standard Keldysh
techniques see e.g. \onlinecite{key-60, key-44, key-82}.) We will
calculate the following correlator: \begin{widetext} \begin{equation}
\left\langle \gamma\left(0\right)\gamma\left(T\right)\right\rangle =\left\langle \gamma\;\left(\widetilde{\mathcal{T}}\; e^{+i\int_{0}^{\mathrm{T}}H\left(\tau\right)\, d\tau}\right)\;\gamma\;\left({\mathcal{T}}\; e^{-i\int_{0}^{\mathrm{T}}H\left(\tau\right)\, d\tau}\right)\right\rangle \;.\label{eq:Tracetocompute}\end{equation}
 Here the expectation value is taken relative to the density matrix
$\rho_{0}$ at $\tau=0$ while ${\mathcal{T}}$ and $\widetilde{\mathcal{T}}$
stand for time ordering and time antiordering respectively. To make
the computations tractable we will assume that $\rho_{0}=\rho_{\mathrm{therm}}\otimes\rho_{\mathrm{Maj}}$.
Here $\rho_{\mathrm{Maj}}$ is any initial density matrix acting on
the subspace of the Majorana modes while $\rho_{\mathrm{therm}}$
is the thermal density matrix for the regular fermion modes.

To compute the correlator in Eq. (\ref{eq:Tracetocompute}), we will
use Eq. (\ref{eq:GenericHamiltonian}) and work in the interaction
picture with respect to the rest of the Hamiltonian ${H}^{\mathrm{bath}}\left(\{{c}_{\mathrm{i}}^{\phantom{\dagger}},{c}_{\mathrm{i}}^{\dagger},{B}_{\mathrm{i}}^{\phantom{\dagger}},{B}_{\mathrm{i}}^{\dagger}\}\right)$.
We will expand the ordered exponentials in powers of $H$ and collect
and contract all the $\gamma$s to eliminate them. In what follows
will show that \begin{equation}
\left\langle \gamma\left(0\right)\gamma\left(T\right)\right\rangle =\left\langle \left(\widetilde{\mathcal{T}}\; e^{-\int_{0}^{\mathrm{T}}\mathrm{O}\left(\tau\right)\, d\tau}\right)\;\left({\mathcal{T}}\; e^{-\int_{0}^{\mathrm{T}}\mathrm{O}\left(\tau\right)\,\mathrm{d}\tau}\right)\right\rangle \equiv\left\langle {\mathcal{T}_{c}}\; e^{-\sum_{a}\int_{0}^{\mathrm{T}}\mathrm{O}\left(\tau^{a}\right)\mathrm{d}\tau^{a}}\right\rangle \;,\label{eq:Z-equiv}\end{equation}
 where $\mathrm{O}\left(\tau\right)=\sum_{\mathrm{i=1}}^{\mathrm{N}}\left(B_{i}\left(\tau\right)c_{i}\left(\tau\right)-B_{i}^{\dagger}\left(\tau\right)c_{i}^{\dagger}\left(\tau\right)\right)$,
and $\mathcal{T}_{\mathrm{c}}$ stands for the Keldysh ordering that
combines the forward and backward propagation, and the index $a\,\mathrm{=\, t,\, b}$
labels the two pieces (forward and backward) of the ordered product.
(Notice though that the operator $\mathrm{O}\left(\tau\right)$ in
the exponential comes with the same sign in the ${\mathcal{T}}$ and
$\widetilde{\mathcal{T}}$ products.)

\begin{figure}[htbp]
\begin{centering}
\includegraphics[scale=0.8]{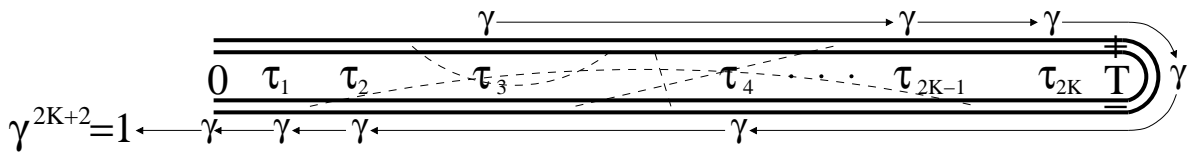} 
\par\end{centering}

\caption{\label{fig:Keldish}The Keldysh contour determining the coherence
of the Majorana zero mode. We consider $2K$ insertions of our interaction
Hamiltonian $\pm i\gamma\sum_{i=1}^{N}\left(B_{i}c_{i}-B_{i}^{\dagger}c_{i}^{\dagger}\right)$
into the Keldysh contour with $\pm$ referring to the forward in time
and backwards in time branches. Several interaction insertions are
shown by dashed lines. To convert this contour to a {}``regular''
Keldysh calculation we commute the Majorana modes ($\gamma$ terms)
including the one at $\tau=\mathrm{T}$ till they are all located
at $\tau=0$ as shown. In the text we describe how to compute commutators
appropriately. }

\end{figure}

\vspace{0.3cm}

Below we give the essential arguments needed to derive Eq.~(\ref{eq:Z-equiv}).
To carry out this program, let us introduce a short-hand notation
$H=\gamma\sum_{i=1}^{N}\left(B_{i}c_{i}-B_{i}^{\dagger}c_{i}^{\dagger}\right)\equiv\gamma\left(\mathbf{B}\mathbf{c}-\mathbf{B}^{\dagger}\mathbf{c}^{\dagger}\right)$.
Now expand Eq.~(\ref{eq:Tracetocompute}) in powers of $H$, and
focus on the term with $N_{\mathrm{b}}+N_{\mathrm{t}}$ insertions,
with $N_{\mathrm{b}}$ from the expansion of the $\widetilde{\mathcal{T}}$-ordered
exponential and $N_{\mathrm{t}}$ from that of the ${\mathcal{T}}$-ordered
exponential. By fermion parity conservation and using our assumption
that the system-bath initial density matrix is factorized we know
that $N_{\mathrm{b}}+N_{\mathrm{t}}=2K$ is even. The insertions of
our interaction Hamiltonian are of the form 
\begin{equation}
\begin{array}{l}
\overbrace{\{\gamma\}}^{\tau=0}\underbrace{\left[i\gamma\left(\mathbf{B}\mathbf{c}-\mathbf{B}^{\dagger}\mathbf{c}^{\dagger}\right)(t_{1}^{b})\right]\cdots\left[i\gamma\left(\mathbf{B}\mathbf{c}-\mathbf{B}^{\dagger}\mathbf{c}^{\dagger}\right)(t_{N_{\mathrm{b}}}^{\mathrm{b}})\right]}_{\text{bottom insertions}}\overbrace{\{\gamma\}}^{\tau=\mathrm{T}}\underbrace{\left[-i\gamma\left(\mathbf{B}\mathbf{c}-\mathbf{B}^{\dagger}\mathbf{c}^{\dagger}\right)(t_{1}^{t})\right]\cdots\left[-i\gamma\left(\mathbf{B}\mathbf{c}-\mathbf{B}^{\dagger}\mathbf{c}^{\dagger}\right)(t_{N_{\mathrm{t}}}^{\mathrm{t}})\right]}_{\text{top insertions}}\;.\end{array}\label{eq:Commutator}\end{equation}
 \vspace{0.3cm}
 \end{widetext} We show in curly brackets $\left\{ \gamma\right\} $
the modes at $\tau=\mathrm{T}$ and at $\tau=0$, to help single them
out for constructing the argument below. Our strategy to convert this
calculation to a {}``regular'' Keldysh calculation will be to move
the Majorana modes ($\gamma$ terms), including the $\{\gamma\}$
at $\tau=\mathrm{T}$, by taking appropriate commutators, till they
are all at the left hand side, adjacent to the $\{\gamma\}$ inserted
at $\tau=0$. We will move along the contour ordering direction (see
Fig.~\ref{fig:Keldish}). We will then use the relation $\gamma^{2K+2}=1$
to eliminate these modes altogether. All that remains is a computation
of the commutators. Because of the form of the Hamiltonian, computing
commutators is equivalent to computing an overall sign for the term
in the expansion. By noting that the Hamiltonian is bosonic we obtain
that the overall sign is only due to the anti-commutation of the $\gamma$'s
with the $c_{i}$ and $c_{i}^{\dagger}$ inside the $\left(\mathbf{B}\mathbf{c}-\mathbf{B}^{\dagger}\mathbf{c}^{\dagger}\right)$
terms. We shall move each $\gamma$ mode to the very left in two steps:
we first move the mode at $\tau=\mathrm{T}$ to the very left towards
$\tau=0$; then we move all the remaining modes there as well.

In the first part of the procedure is to obtain the contribution of
the Majorana fermion inserted at $\tau=\mathrm{T}$. We note that
the number of $-1$ signs it picks up depends on its position along
the contour relative to the other modes it picks up one $-1$ sign
for very mode it passes so there is an overall sign of $\left(-1\right)^{N_{\mathrm{b}}}$.

Now for the rest working from left to right, the first Majorana mode
that needs to be moved picks up no $-1$ signs as it does not pass
over a $\left(\mathbf{B}\mathbf{c}-\mathbf{B}^{\dagger}\mathbf{c}^{\dagger}\right)$
term, but the second picks up one $-1$ sign as it passes over one
such term. Similarly, the third picks up two $(-1)$ signs, and so
forth. Finally the $2K$th Majorana mode (last to be moved, sitting
all the way to the right) picks up $2K-1$ factors of $-1$. The product
of these factors yields $\left(-1\right)^{K\left(2K-1\right)}=\left(-1\right)^{K}=\left(-i\right)^{N_{\mathrm{t}}+N_{\mathrm{b}}}$.

\begin{widetext} Thus eliminating the $\gamma$'s in Eq.~(\ref{eq:Commutator})
leads to an overall sign ${(-i)^{N_{\mathrm{b}}+N_{\mathrm{t}}}\times(-1)^{N_{\mathrm{b}}}}$,
which then allows us to replace terms of the form Eq.~(\ref{eq:Commutator})
by \begin{equation}
\begin{array}{l}
\underbrace{\left[-\left(\mathbf{B}\mathbf{c}-\mathbf{B}^{\dagger}\mathbf{c}^{\dagger}\right)(t_{1}^{\mathrm{b}})\right]\cdots\left[-\left(\mathbf{B}\mathbf{c}-\mathbf{B}^{\dagger}\mathbf{c}^{\dagger}\right)(t_{N_{\mathrm{b}}}^{\mathrm{b}})\right]}_{\text{bottom insertions}}\underbrace{\left[-\left(\mathbf{B}\mathbf{c}-\mathbf{B}^{\dagger}\mathbf{c}^{\dagger}\right)(t_{1}^{\mathrm{t}})\right]\cdots\left[-\left(\mathbf{B}\mathbf{c}-\mathbf{B}^{\dagger}\mathbf{c}^{\dagger}\right)(t_{N_{\mathrm{t}}}^{\mathrm{t}})\right]}_{\text{top insertions}}\;.\end{array}\label{eq:Commutator-Z}\end{equation}
 These are precisely the terms that appear in the series expansion
of Eq.~(\ref{eq:Z-equiv}), and therefore we can continue the calculation
utilizing this expression. We should point out that for complex fermions
coming from Majorana insertion ${\mathcal{T}}_{c}$ corresponds to
literal ordering on the Keldysh contour, without any fermionic minus
signs, because the original Hamiltonian was bosonic {[}this can also
be seen step-by-step in going from Eq.~(\ref{eq:Commutator}) to
Eq.~(\ref{eq:Commutator-Z}){]}. This fact leads to the modified
sign for the fermionic ${\mathcal{T}}_{c}$-ordering: \begin{equation}
\mathcal{T}_{c}\left[c_{i}^{\dagger}\left(t_{1}\right)c_{i}\left(t_{2}\right)\right]\equiv\left\{ \begin{array}{llll}
\theta\left(t_{1}-t_{2}\right)\; c_{i}^{\dagger}\left(t_{1}\right)c_{i}\left(t_{2}\right)+\theta\left(t_{2}-t_{1}\right)\; c_{i}\left(t_{2}\right)c_{i}^{\dagger}\left(t_{1}\right) & ,\; t_{1},t_{2}\;\mathrm{on\; top}\\
c_{i}^{\dagger}\left(t_{1}\right)c_{i}\left(t_{2}\right) & ,\; t_{1}\;\mathrm{on\; bottom},t_{2}\;\mathrm{on\; top}\\
c_{i}\left(t_{2}\right)c_{i}^{\dagger}\left(t_{1}\right) & ,\; t_{1}\;\mathrm{on\; top},t_{2}\;\mathrm{on\; bottom}\\
\theta\left(t_{2}-t_{1}\right)\; c_{i}^{\dagger}\left(t_{1}\right)c_{i}\left(t_{2}\right)+\theta\left(t_{1}-t_{2}\right)\; c_{i}\left(t_{2}\right)c_{i}^{\dagger}\left(t_{1}\right) & ,\; t_{1},t_{2}\;\mathrm{on\; bottom}\;.\end{array}\right.\label{eq:WeirdTimeordering}\end{equation}
 \vspace{0.3cm}

Now, we turn our attention to the computation of Eq.~(\ref{eq:Z-equiv}).
We do so in steps, computing the expectation values by first tracing
the fermions ($c_{i}\,,\, c_{i}^{\dagger}$) and then subsequently
tracing the bosonic degrees of freedom. Even in the case where there
are interactions for the fermions, we can still treat the theory as
quadratic in the fermions and include the interactions (with photons
or phonons) as a coupling of the fermionic bilinears with the mediating
bosons, which we label by $\phi$. Alternatively, we may think of
the fields $\phi$ fields as Hubbard-Stratonovich decoupling fields\cite{key-78}.

We can thus write \begin{eqnarray}
 &  & \left\langle {\mathcal{T}_{\mathrm{c}}}\; e^{-\sum_{a}\int_{0}^{\mathrm{T}}\left(\mathbf{B}\mathbf{c}-\mathbf{B}^{\dagger}\mathbf{c}^{\dagger}\right)\left(\tau^{a}\right)\;\mathrm{d}\tau^{a}}\right\rangle ={\mathcal{Z}}^{-1}\;\int\left(\prod_{a}\mathcal{D}\mathbf{B}_{a}\,\mathcal{D}\mathbf{B}_{a}^{\dagger}\right)\; e^{i{\mathcal{S}}_{\mathbf{B}}[\mathbf{B}_{a}\,\mathbf{B}_{a}^{\dagger}]}\;\int\left(\prod_{a}\mathcal{D}\mathbf{\phi}_{a}\,\mathcal{D}\mathbf{\phi}_{a}^{\dagger}\right)\; e^{i{\mathcal{S}}_{\mathbf{\phi}}[\mathbf{\phi}_{a}\,\mathbf{\phi}_{a}^{\dagger}]}\nonumber \\
 &  & \qquad\qquad\qquad\qquad\qquad\times\;\exp\left(\frac{1}{2}\,{\sum_{a,b}\;\int_{0}^{\mathrm{T}}\!\!\mathrm{d}\tau_{1}^{a}\int_{0}^{\mathrm{T}}\!\!\mathrm{d}\tau_{2}^{b}\;\;\left\langle {\mathcal{T}_{c}}\left[\left(\mathbf{B}\mathbf{c}-\mathbf{B}^{\dagger}\mathbf{c}^{\dagger}\right)\left(\tau_{1}^{a}\right)\;\left(\mathbf{B}\mathbf{c}-\mathbf{B}^{\dagger}\mathbf{c}^{\dagger}\right)\left(\tau_{2}^{b}\right)\right]\right\rangle _{\mathbf{c},\mathbf{c}^{\dagger}}}\right)\;.\end{eqnarray}
 We remind the reader that all functional integrals are along the
Keldysh contour. The action ${\mathcal{S}}_{\mathbf{\phi}}$ is that
of the interaction mediator field $\phi$ and contains the dressing
from the integration of the fermions, which are integrated out first
as explained above. The normalization ${\mathcal{Z}}$ is \begin{eqnarray}
{\mathcal{Z}}=\int\left(\prod_{a}\mathcal{D}\mathbf{B}_{a}\,\mathcal{D}\mathbf{B}_{a}^{\dagger}\right)\; e^{i{\mathcal{S}}_{\mathbf{B}}[\mathbf{B}_{a}\,\mathbf{B}_{a}^{\dagger}]}\;\int\left(\prod_{a}\mathcal{D}\mathbf{\phi}_{a}\,\mathcal{D}\mathbf{\phi}_{a}^{\dagger}\right)\; e^{i{\mathcal{S}}_{\mathbf{\phi}}[\mathbf{\phi}_{a}\,\mathbf{\phi}_{a}^{\dagger}]}\;.\label{eq:nonm-partial-fermion}\end{eqnarray}

This procedure works because it possible to calculate partition functions,
Green's functions, integrate fields out etc. along any contour, in
particular along the Keldysh contour as used here. We then express
the fermionic correlators in terms of their Green's function, \begin{eqnarray}
\left\langle {\mathcal{T}_{\mathrm{c}}}\left[\left(\mathbf{B}\mathbf{c}-\mathbf{B}^{\dagger}\mathbf{c}^{\dagger}\right)\left(\tau_{1}^{a}\right)\;\left(\mathbf{B}\mathbf{c}-\mathbf{B}^{\dagger}\mathbf{c}^{\dagger}\right)\left(\tau_{2}^{b}\right)\right]\right\rangle _{\mathbf{c},\mathbf{c}^{\dagger}} & = & -{B_{i}}\left(\tau_{1}^{a}\right){B_{j}}^{\dagger}\left(\tau_{2}^{b}\right)\;\left\langle {\mathcal{T}_{\mathrm{c}}}\left[{c_{i}}\left(\tau_{1}^{a}\right)\;{c_{j}}^{\dagger}\left(\tau_{2}^{b}\right)\right]\right\rangle \nonumber \\
 &  & -\;{B_{i}}^{\dagger}\left(\tau_{1}^{a}\right){B_{j}}\left(\tau_{2}^{b}\right)\;\left\langle {\mathcal{T}_{\mathrm{c}}}\left[{c_{i}}^{\dagger}\left(\tau_{1}^{a}\right)\;{c_{j}}\left(\tau_{2}^{b}\right)\right]\right\rangle \nonumber \\
 & \equiv & -\mathbf{B}\left(\tau_{1}^{a}\right)\; G_{\mathrm{F,e}}^{\phi}\left(\tau_{1}^{a},\tau_{2}^{b}\right)\;\mathbf{B}^{\dagger}\left(\tau_{2}^{b}\right)\nonumber \\
 &  & -\;\mathbf{B}^{\dagger}\left(\tau_{1}^{a}\right)\; G_{\mathrm{F,h}}^{\phi}\left(\tau_{1}^{a},\tau_{2}^{b}\right)\;\mathbf{B}\left(\tau_{2}^{b}\right),\end{eqnarray}
 where the $G_{\mathrm{F,e}}^{\phi}\left(\tau_{1}^{a},\tau_{2}^{b}\right)$
and $G_{\mathrm{F,h}}^{\phi}\left(\tau_{1}^{a},\tau_{2}^{b}\right)$
are, respectively, the electron and hole fermionic Green's function,
and we have used the fact that the bosonic fields $\mathbf{B}\,,\,\mathbf{B}^{\dagger}$
can be treated as c-numbers as they are inside the bosonic path integral.
As stated previously $G_{\mathrm{F,e}}^{\phi}\left(\tau_{1}^{a},\tau_{2}^{b}\right)$
and $G_{\mathrm{F,h}}^{\phi}\left(\tau_{1}^{a},\tau_{2}^{b}\right)$
are slightly unusual Green's functions, with no fermionic minus signs
(only plus signs), as shown in Eq.~(\ref{eq:WeirdTimeordering}).
Let us define $D_{\mathrm{F}}^{\phi}\left(\tau_{1}^{a},\tau_{2}^{b}\right)=\; G_{\mathrm{F,h}}^{\phi}\left(\tau_{1}^{a},\tau_{2}^{b}\right)+\; G_{\mathrm{F,e}}^{\phi}\left(\tau_{2}^{b},\tau_{1}^{a}\right)$,
so we can then write \begin{eqnarray}
 &  & \left\langle \gamma\left(0\right)\gamma\left(T\right)\right\rangle ={\mathcal{Z}}^{-1}\;\int\left(\prod_{a}\mathcal{D}\mathbf{B}_{a}\,\mathcal{D}\mathbf{B}_{a}^{\dagger}\right)\; e^{i{\mathcal{S}}_{\mathbf{B}}[\mathbf{B}_{a}\,\mathbf{B}_{a}^{\dagger}]}\;\int\left(\prod_{a}\mathcal{D}\mathbf{\phi}_{a}\,\mathcal{D}\mathbf{\phi}_{a}^{\dagger}\right)\; e^{i{\mathcal{S}}[\mathbf{\phi}_{a}\,\mathbf{\phi}_{a}^{\dagger}]}\nonumber \\
 &  & \qquad\qquad\qquad\times\;\exp\left(-\frac{1}{2}\,{\sum_{a,b}\;\int_{0}^{\mathrm{T}}\!\!\mathrm{d}\tau_{1}^{a}\int_{0}^{\mathrm{T}}\!\!\mathrm{d}\tau_{2}^{b}\;\;\mathbf{B}^{\dagger}\left(\tau_{1}^{a}\right)\; D_{\mathrm{F}}^{\phi}\left(\tau_{1}^{a},\tau_{2}^{b}\right)\;\mathbf{B}\left(\tau_{2}^{b}\right)}\right)\;.\label{eq:FermionExpectationPath}\end{eqnarray}
 We remark that the expression in Eq.~(\ref{eq:FermionExpectationPath})
was derived without any approximations. It holds for interacting electrons
as well, as long as the interactions are included via an external
bosonic field denoted by $\phi$ above. Furthermore we would like
to note that though it is not used anywhere in this paper, but a similar
path integral formulation using Grassmann variables may be done without
any decoupling fields, for regular quartic $\sim\Psi^{\dagger}\left(\vec{x}\right)\Psi^{\dagger}\left(\vec{x}\right)\Psi\left(\vec{x}\right)\Psi\left(\vec{x}\right)$
fermionic interactions. A systematic Keldysh diagrammatic perturbation
theory may be derived from it.

\vspace{0.3cm}

For future use we note that to compute the coherence of a Majorana
mode it is often enough to compute the four diagrams shown in Fig.
(\ref{fig:MajoranaPhonons}). Following Eq. (\ref{eq:FermionExpectationPath}),
their sum may be explicitly written as:\begin{eqnarray}
V\left(T\right) & \equiv & \sum_{a,b}\;\int_{0}^{\mathrm{T}}\!\!\mathrm{d}\tau_{1}^{a}\int_{0}^{\mathrm{T}}\!\!\mathrm{d}\tau_{2}^{b}\;\;\mathbf{B}^{\dagger}\left(\tau_{1}^{a}\right)\; D_{\mathrm{F}}^{\phi}\left(\tau_{1}^{a},\tau_{2}^{b}\right)\;\mathbf{B}\left(\tau_{2}^{b}\right)\nonumber \\
 & = & 2\sum_{i}\left\{ \int_{0}^{\mathrm{T}}d\tau_{1}\int_{0}^{\mathrm{T}}d\tau_{2}\left[\mathcal{T}\left({B_{i}}^{\dagger}\left(\tau_{1}^{\mathrm{t}}\right){B_{i}}\left(\tau_{2}^{\mathrm{\mathrm{t}}}\right)\right)\times\left(\theta\left(\tau_{1}-\tau_{2}\right)\left\langle c_{i}^{\dagger}\left(\tau_{1}\right)c_{i}\left(\tau_{2}\right)\right\rangle +\theta\left(\tau_{2}-\tau_{1}\right)\left\langle c_{i}\left(\tau_{2}\right)c_{i}^{\dagger}\left(\tau_{1}\right)\right\rangle \right)\right.\right.\nonumber \\
 & + & \widetilde{\mathcal{T}}\left({B_{i}}^{\dagger}\left(\tau_{1}^{\mathrm{b}}\right){B_{i}}\left(\tau_{2}^{\mathrm{\mathrm{b}}}\right)\right)\times\left(\theta\left(\tau_{2}-\tau_{1}\right)\left\langle c_{i}^{\dagger}\left(\tau_{1}\right)c_{i}\left(\tau_{2}\right)\right\rangle +\theta\left(\tau_{1}-\tau_{2}\right)\left\langle c_{i}\left(\tau_{2}\right)c_{i}^{\dagger}\left(\tau_{1}\right)\right\rangle \right)\label{eq:TimeOrderedEquations}\\
 & + & \left.\left.\left({B_{i}}^{\dagger}\left(\tau_{1}^{\mathrm{t}}\right){B_{i}}\left(\tau_{2}^{\mathrm{\mathrm{b}}}\right)\left\langle c_{i}^{\dagger}\left(\tau_{1}\right)c_{i}\left(\tau_{2}\right)\right\rangle +{B_{i}}\left(\tau_{1}^{\mathrm{t}}\right){B_{i}^{\dagger}}\left(\tau_{2}^{\mathrm{\mathrm{b}}}\right)\left\langle c_{i}\left(\tau_{1}\right)c_{i}^{\dagger}\left(\tau_{2}\right)\right\rangle \right)\right]\right\} \nonumber \end{eqnarray}
Here $\mathcal{T}$, $\widetilde{\mathcal{T}}$ refer to time ordering
and time anti-ordering operators. This form places the time ordering
or antiordering terms ($\mathcal{T}\left({B_{i}}^{\dagger}\left(\tau_{1}^{\mathrm{t}}\right){B_{i}}\left(\tau_{2}^{\mathrm{\mathrm{t}}}\right)\right)$)
with the appropriate fermion correlators so it can be used directly
in calculations without having to use a path integral. The factor
of two going from the first to the second line comes from a symmetry
$\tau_{1}\leftrightarrow\tau_{2}$ (which also allowed us to simplify
Eq. (\ref{eq:TimeOrderedEquations}) above to contain six rather then
twelve terms). Because of exponentiation of disconnected diagrams,
if we can safely ignore higher order correlations among the ${B_{i}}$'s,
we may write that:\begin{equation}
\left\langle \gamma\left(0\right)\gamma\left(T\right)\right\rangle =\mathrm{e}^{-\frac{1}{2}\left\langle V\left(T\right)\right\rangle }\;.\label{eq:master-corgeneralr}\end{equation}

A quick way to derive the extra factor of $\frac{1}{2}$ in Eq. (\ref{eq:master-corgeneralr})
above is by noting that it is a symmetry factor associated with the
ability to permute the two Majorana insertions without changing the
diagram {[}alternatively we can do a combinatorial check, or use Eq.
(\ref{eq:FermionExpectationPath}){]}.

Let us illustrate with a few simple examples how one can use the expression
for the Majorana correlations $\left\langle \gamma\left(0\right)\gamma\left(T\right)\right\rangle $
in Eq.~(\ref{eq:FermionExpectationPath}) to calculate the the decay
rates of topological memories. We then deploy this expression in detailed
studies for fluctuating Hamiltonians in Section~\ref{sec:Fluctuating-Hamiltonians}.

\begin{figure}[htbp]
\begin{centering}
\includegraphics[scale=0.65]{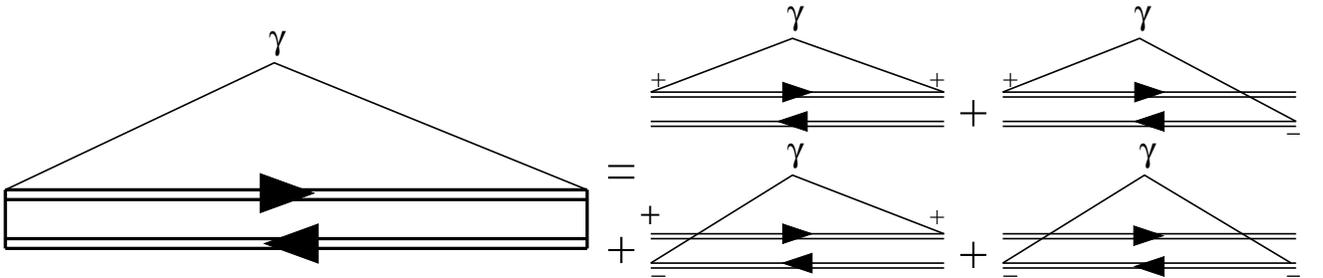} 
\par\end{centering}

\caption{\label{fig:MajoranaPhonons}The four diagrams relevant to calculating
$V\left(T\right)$ in the main text. We need to sum over four possible
orderings of the Majorana insertions on the Keldysh contour. The value
is given by a sum of terms like ${B_{i}}^{\dagger}\left(\tau_{1}^{\mathrm{t}}\right){B_{i}}\left(\tau_{2}^{\mathrm{\mathrm{b}}}\right)\left\langle c_{i}^{\dagger}\left(\tau_{1}\right)c_{i}\left(\tau_{2}\right)\right\rangle $.}

\end{figure}

\subsection{\label{sub:Simple-examples}Simple examples}

\label{sec:Keldysh-examples}

Let us consider simple cases where the $B_{i}$ are simply 
constants $\Gamma_{i}$, switched on at $\tau=0$. In this case the
expression in Eq.~(\ref{eq:FermionExpectationPath}) simplifies to
\begin{eqnarray}
 &  & \left\langle \gamma\left(0\right)\gamma\left(T\right)\right\rangle ={\mathcal{Z}}^{-1}\;\;\int\left(\prod_{a}\mathcal{D}\mathbf{\phi}_{a}\,\mathcal{D}\mathbf{\phi}_{a}^{\dagger}\right)\; e^{i{\mathcal{S}}[\mathbf{\phi}_{a}\,\mathbf{\phi}_{a}^{\dagger}]}\nonumber \\
 &  & \qquad\qquad\qquad\times\;\exp\left(-\frac{1}{2}\,{\sum_{a,b}\;\int_{0}^{\mathrm{T}}\!\!\mathrm{d}\tau_{1}^{a}\int_{0}^{\mathrm{T}}\!\!\mathrm{d}\tau_{2}^{b}\;\;\mathbf{\Gamma}^{\dagger}\; D_{\mathrm{F}}^{\phi}\left(\tau_{1}^{a},\tau_{2}^{b}\right)\;\mathbf{\Gamma}}\right)\nonumber \\
 &  & \qquad\qquad\qquad=\exp\left(-\frac{1}{2}\,{\sum_{a,b}\;\int_{0}^{\mathrm{T}}\!\!\mathrm{d}\tau_{1}^{a}\int_{0}^{\mathrm{T}}\!\!\mathrm{d}\tau_{2}^{b}\;\;\mathbf{\Gamma}^{\dagger}\;\overline{D}_{F}^{(2)}\left(\tau_{1}^{a},\tau_{2}^{b}\right)\;\mathbf{\Gamma}}\;+\;\dots\right)\;,\label{eq:FinalKeldish}\end{eqnarray}
 where $\overline{\mathrm{D}}_{\mathrm{F}}^{(2)}\left(\tau_{1}^{a},\tau_{2}^{b}\right)=\;\overline{\mathrm{G}}_{\mathrm{F,h}}^{(2)}\left(\tau_{1}^{a},\tau_{2}^{b}\right)+\;\overline{\mathrm{G}}_{\mathrm{F,e}}^{(2)}\left(\tau_{2}^{b},\tau_{1}^{a}\right)$,
with $\overline{\mathrm{G}}_{\mathrm{F,h}}^{(2)}$ and $\overline{\mathrm{G}}_{\mathrm{F,e}}^{(2)}$
exact 2-point electron and hole Keldysh propagators, including the
effects of interactions. To be explicit at this level of approximation
our formalism handles all the dynamics of the $\phi_{a}$ fields but
treats fermionic interactions to quadratic order. The $\dots$ stand
for terms of order ${\mathcal{O}}(\mathbf{\Gamma}^{4})$ that involve
the 4-point Green's functions $\overline{\mathrm{G}}^{(4)}$. We shall
not do so in this paper, but by including these ${\mathcal{O}}(\mathbf{\Gamma}^{4})$
and higher terms it is possible to handle all fermionic interactions
as well.

Taking into account all the four cases in the sum over top and bottom
insertions $\sum_{a,b}$, one can write \begin{eqnarray}
\frac{1}{2}\,{\sum_{a,b}\;\int_{0}^{\mathrm{T}}\!\!\!\!\mathrm{d}\tau_{1}^{a}\int_{0}^{\mathrm{T}}\!\!\!\!\mathrm{d}\tau_{2}^{b}\;\;\mathbf{\Gamma}^{\dagger}\;\overline{\mathrm{D}}_{\mathrm{F}}^{(2)}\left(\tau_{1}^{a},\tau_{2}^{b}\right)\;\mathbf{\Gamma}}=\sum_{i,j}\!\!\int_{0}^{\mathrm{T}}\!\!\!\!\mathrm{d}\tau_{1}\int_{0}^{\mathrm{T}}\!\!\!\!\mathrm{d}\tau_{2}\;\;\Gamma_{i}^{*}\left(\langle\{c_{i}^{\dagger}(\tau_{1}),c_{j}(\tau_{2})\}\rangle\right)\Gamma_{j}\;.\label{eq:partial-fermion-static}\end{eqnarray}
 \end{widetext} We now consider a case where this formula will be
particularly useful. We Consider the case when the bath is described
by the Hamiltonian \begin{equation}
H=\gamma\sum_{i=1}^{N}\left(\Gamma_{i}c_{i}-\Gamma_{i}^{*}c_{i}^{\dagger}\right)+\sum_{i=1}^{N}\epsilon_{i}c_{i}^{\dagger}c_{i}\;.\label{eq:Main Hamiltonian}\end{equation}
 In this case we have \begin{equation}
\langle\{c_{i}^{\dagger}(\tau_{1}),c_{j}(\tau_{2})\}\rangle=\delta_{ij}\; e^{-i\epsilon_{\mathrm{i}}(\tau_{1}-\tau_{2})}\end{equation}
 with $\epsilon_{i}$ the energy of mode $i$. It follows by substitution
in Eq.~(\ref{eq:partial-fermion-static}) and then in Eq.~(\ref{eq:FermionExpectationPath})
that \begin{equation}
\left\langle \gamma\left(0\right)\gamma\left(\mathrm{T}\right)\right\rangle =\mathrm{e}^{-2\sum_{i}|\Gamma_{i}|^{2}|\int_{0}^{\mathrm{T}}d\tau\; e^{-i\mathrm{\epsilon}_{i}\tau}|^{2}}\;,\label{eq:master-corr}\end{equation}
 or \begin{equation}
\left\langle \gamma\left(0\right)\gamma\left(\mathrm{T}\right)\right\rangle =\mathrm{e}^{-4\sum_{i}\frac{|\Gamma_{i}|^{2}}{\epsilon_{i}^{2}}[1-\cos(\epsilon_{i}T)]}\;.\label{eq:non-int-static}\end{equation}
 If the bath has energy eigenenergies $\epsilon_{i}$ away from zero
energy (\textit{i.e.}, there is a gap $\tilde{\epsilon}<|\epsilon_{i}|$),
we may drop the oscillating terms in the limit of $\mathrm{T}\gg1/\tilde{\epsilon}$,
so we can write \begin{equation}
\left\langle \gamma\left(0\right)\gamma\left(\mathrm{T}\right)\right\rangle \approx e^{-4\sum_{i}\frac{|\Gamma_{i}|^{2}}{\epsilon_{i}^{2}}}\;,\quad\mathrm{T}\gg1/\tilde{\epsilon}\;.\label{eq:LongTimeStatic}\end{equation}
 In this case, the Majorana memory decays to $\mathrm{T}$ independent
plateaus at large times. Thus, as long as the sum $\sum\frac{|\Gamma_{i}|^{2}}{\epsilon_{i}^{2}}$
converges, the memory is retained to a finite extent. This result
is confirmed by a time-independent re-diagonalization in the presence
of the $\Gamma_{i}$, which is shown explicitly in Appendix~\ref{sec:QuadraticHamiltonian}
where a new exact zero mode is calculated. Here we simply note that
the finite depletion found in this case is a simple consequence of
the fact that the modes change once the coupling is switched on. Also,
we compute the sum $\sum\frac{|\Gamma_{i}|^{2}}{\epsilon_{i}^{2}}$,
and find it to be finite, for a specific tunneling model in Appendix~\ref{sub:Summation-of-Eq.SumBounds}.

\vspace{0.3cm}


%
\begin{figure}[htbp]
\begin{centering}
\includegraphics[scale=1.05]{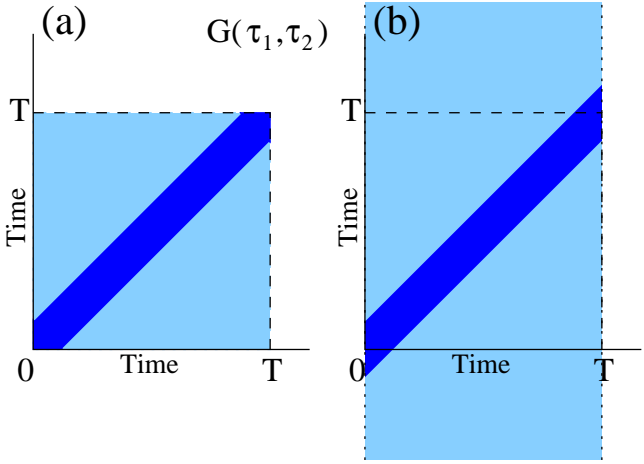} 
\par\end{centering}

\caption{\label{fig:Correlator}The two time correlators of the tunneling amplitude
$G_{\Gamma}\left(\tau_{1},\tau_{2}\right)=\left\langle \Gamma^{*}\left(\tau_{1}\right)\Gamma\left(\tau_{2}\right)\right\rangle $.
a) The shaded region represents the actual area of integration for
Eq. (\ref{eq:CorrelatorMatrix}). The darker stripe represents the
area of large values for the correlator. This represents strong correlations
in the tunneling amplitudes. From this we see that the majority of
the integrals appearing in Eq. (\ref{eq:CorrelatorMatrix}) come from
times when $\tau_{\mathrm{1}}\cong\tau_{2}$. b) A simplified integration
area. The darkly shaded area of large correlators does not change
significantly. As such geometrically we see that this should not change
the values of the various correlation functions we are studying. From
this it is particularly easy to derive the estimates used in Eq. (\ref{eq:LongTimeCorrelatorMatrix}),
in particular the linear in $\mathrm{T}$ scaling can now be derived
by simply changing co-ordinates in the integral in Eq. (\ref{eq:CorrelatorMatrix}). }

\end{figure}

\section{Fluctuating Hamiltonians \label{sec:Fluctuating-Hamiltonians}}

So far we have studied static Hamiltonians. To gain further insight
it is interesting to extend our results to fluctuating couplings (which
may come from time dependent classical fluctuations or from quantum
dynamics). We shall focus on three cases, in all three the fermionic
action is quadratic. In the first case we study we consider the situation
when the $B_{i}$ are simply replaced by classical variables $\Gamma_{i}$,
like we did in Sec.~\ref{sec:Keldysh-examples}, but now they depend
on time. The second case is that when the energies $\epsilon_{i}$
of the electrons in the bath fluctuate in time, because of environmental
fluctuations. The third case is a generalization of the first one,
where we treat the $B_{i}$ quantum mechanically with their fluctuations
governed by a quadratic action. We treat the first two cases here,
and the third, more technical one, in Appendix~\ref{sub:Quantum-Fluctuations}.

In the first two cases, one can generalize the expression in Eq.~(\ref{eq:master-corr})
simply by taking $\Gamma_{i}\to\Gamma_{i}(\tau)$ or $\epsilon_{i}\to\epsilon_{i}(\tau)$:
\begin{eqnarray}
\left\langle \gamma\left(0\right)\gamma\left(T\right)\right\rangle  & = & \mathrm{e}^{{-2\sum_{i}\left|\int_{0}^{\mathrm{T}}d\tau\;\Gamma_{i}(\tau)\; e^{-i\int_{0}^{\tau}dt\,\epsilon_{i}(t)}\right|^{2}}}\nonumber \\
 & = & \prod_{i}\mathrm{e}^{-2\left|\int_{0}^{\mathrm{T}}\mathrm{d}\tau\;\Gamma_{i}(\tau)\; e^{-i\int_{0}^{\tau}dt\,\epsilon_{i}(t)}\right|^{2}}\,,\label{eq:master-corr-t-dep}\end{eqnarray}
 and then average over statistical fluctuations of the $\Gamma_{i}(\tau)$
and $\epsilon_{i}(\tau)$. 

The computation of the Majorana correlations can be greatly simplified
as follows. Notice that, for each mode $i$, the argument in the exponential
in Eq.~(\ref{eq:master-corr-t-dep}) can be viewed as the magnitude
square of the position $\vec{Z}_{i}$ of a particle moving in two-dimensions,
or alternatively the modulus square of a complex number $\mathrm{Z}_{i}$
moving on the plane: \begin{equation}
Z_{i}(\mathrm{T})=\sqrt{2}\int_{0}^{\mathrm{T}}d\tau\;\Gamma_{i}(\tau)\;\mathrm{e}^{-\mathrm{i}\int_{0}^{\tau}dt\,\epsilon_{i}\left(t\right)}\;,\label{eq:Displacement}\end{equation}
 with \begin{equation}
\left\langle \gamma\left(0\right)\gamma\left(T\right)\right\rangle =\prod_{\mathrm{i}}\mathrm{e}^{-|\vec{Z}_{i}|^{2}}\;.\label{eq:ExpectationValueDisplacement}\end{equation}
 Below we will argue both in the cases of fluctuating amplitudes $\Gamma_{i}(\tau)$
and energies $\epsilon_{i}(\tau)$ that the probability distribution
for the {}``position'' $\vec{Z}_{i}$ is Gaussian: \begin{equation}
\mathrm{P}(\vec{{Z}}_{i})=\frac{1}{2\pi\sigma_{i}^{2}(\mathrm{T})}\;\exp\left(-\frac{1}{2}\frac{|\vec{Z}_{i}|^{2}}{\sigma_{i}^{2}(\mathrm{T})}\right)\;,\label{eq:AnsweExpectationValueDisplacement}\end{equation}
 with $\sigma_{i}(\mathrm{T})$ the time-dependent width of the distribution,
which we will compute below for each case. With this Gaussian distribution
for the $\vec{Z}_{i}$, we can compute the average Majorana correlation,
\begin{eqnarray}
\overline{\left\langle \gamma\left(0\right)\gamma\left(\mathrm{T}\right)\right\rangle } & = & \prod_{i}\int d^{2}Z_{i}\; P(\vec{Z}_{i})\; e^{-|\vec{Z}_{i}|^{2}}\nonumber \\
 & = & \prod_{i}\left[1+2\,\sigma_{i}^{2}(\mathrm{T})\right]^{-1}\nonumber \\
 & \approx & \exp\left[-2\,\sum_{i}\sigma_{i}^{2}(\mathrm{T})\right]\;.\label{eq:FinalSimpleCorrelator}\end{eqnarray}
 In the last step we assumed that there are many modes in the fermionic
bath, each making a small contribution (or order inverse volume) so
we may re-exponentiate the product. The examples below are studied
using this expression.

\subsection{\label{sub:Fluctuating-amplitudes}Fluctuating amplitudes}

The fluctuations of the $\Gamma_{i}$ are assumed to be Gaussian distributed
according to \begin{widetext} \begin{equation}
\mathrm{P}\left(\{\Gamma_{\mathrm{i}}(\tau),\Gamma_{\mathrm{i}}^{*}(\tau)\}\right)={\mathcal{N}}^{-1}\; e^{-\frac{1}{2}\int_{-\infty}^{\infty}d\tau_{1}\int_{-\infty}^{\infty}d\tau_{2}\;\Gamma_{\mathrm{i}}^{*}(\tau_{1})\; G_{\Gamma_{i}}^{-1}(\tau_{1},\tau_{2})\;\Gamma_{\mathrm{i}}(\tau_{2})}\;.\label{eq:SimpleProbabilityDistribution}\end{equation}
 Let us show that the distribution of the $\mathrm{P}(\vec{Z}_{i})$
is Gaussian, and relate $\sigma_{i}(\mathrm{T})$ to the fluctuations
of the $\Gamma_{i}$. That the distribution $\mathrm{P}(\vec{Z}_{i})$
should be Gaussian is not surprising since at long times the particle
is diffusing. We can write for the characteristic function distribution
(Fourier transform of the probability distribution $\mathrm{P}(\vec{Z}_{i})$);
\begin{eqnarray}
\tilde{P}\left(\vec{k}\right) & = & \int d^{2}\vec{\mathrm{Z}}_{i}\; P(\vec{\mathrm{Z}}_{i})\; e^{-i\vec{k}\cdot\vec{Z}_{i}}\nonumber \\
 & = & {\mathcal{N}}^{-1}\int\mathcal{D}\Gamma_{i}\,\mathcal{D}\Gamma_{i}^{*}\;\;\mathrm{e}^{-\frac{1}{2}\int_{-\infty}^{\infty}d\tau_{1}\int_{-\infty}^{\infty}d\tau_{2}\;\Gamma_{i}^{*}(\tau_{1})\; G_{\Gamma_{i}}^{-1}(\tau_{1},\tau_{2})\;\Gamma_{i}(\tau_{2})}\nonumber \\
 &  & \qquad\qquad\qquad\times e^{-i\frac{1}{2}\, k^{*}\sqrt{2}\int_{0}^{\mathrm{T}}d\tau\;\Gamma_{i}(\tau)\; e^{-i\epsilon_{i}\tau}}\;\times e^{-i\frac{1}{2}\, k\sqrt{2}\int_{0}^{\mathrm{T}}d\tau\;\Gamma_{i}^{*}(\tau)\; e^{+i\epsilon_{i}\tau}}\nonumber \\
 & = & \exp\left(-\frac{1}{2}\,|k|^{2}\times2\times\int_{0}^{\mathrm{T}}\!\!\!\! d\tau_{1}\;\int_{0}^{\mathrm{T}}\!\!\!\! d\tau_{2}\;\; e^{-i\epsilon_{i}\tau_{1}}\; G_{\Gamma_{i}}(\tau_{1},\tau_{2})\; e^{+i\epsilon_{i}\tau_{2}}\right)\;.\label{eq:CharacteristicFunction}\end{eqnarray}
 Therefore, the distribution $P\left(\vec{Z_{i}}\right)$ is Gaussian,
with a variance given by \begin{equation}
\sigma_{i}^{2}(\mathrm{T})=2\int_{0}^{\mathrm{T}}\!\!\!\! d\tau_{1}\;\int_{0}^{\mathrm{T}}\!\!\!\! d\tau_{2}\;\; e^{-i\epsilon_{i}\tau_{1}}\; G_{\Gamma_{i}}(\tau_{1},\tau_{2})\; e^{+i\epsilon_{i}\tau_{2}}\;.\label{eq:CorrelatorMatrix}\end{equation}

If the noise correlations are invariant under time-translation, then
$G_{\Gamma_{i}}(\tau_{1},\tau_{2})=G_{\Gamma_{i}}(\tau_{1}-\tau_{2})$.
We can expand these correlations in frequency domain, $G_{\Gamma_{i}}(\tau_{1}-\tau_{2})=\int_{-\infty}^{\infty}d\omega\;\tilde{G}_{\Gamma_{i}}(\omega)\; e^{-i\omega(\tau_{1}-\tau_{2})}$.

We proceed to compute $\sigma_{i}^{2}(\mathrm{T})$ in Eq.~(\ref{eq:CorrelatorMatrix})
for two distinct cases of low and of high frequency noise.

\vspace{0.3cm}
 \end{widetext}

\subsubsection*{Case I: Low-frequency noise}

In this case, we shall assume that all frequencies $\omega$ for which
$\tilde{G}_{\Gamma_{i}}(\omega)$ has significant weight fall below
the fermionic energies $\epsilon_{i}$. It the follows that \begin{eqnarray}
\sigma_{i}^{2}(\mathrm{T}) & = & 2\int_{|\omega|\ll\tilde{\epsilon}}\!\!\!\!\!\! d\omega\;\sum_{i}\frac{1-\cos[(\epsilon_{i}+\omega)T]}{(\epsilon_{i}+\omega)^{2}}\;\tilde{G}_{\Gamma_{i}}(\omega)\;\nonumber \\
 & \approx & 2\sum_{i}\frac{1}{\epsilon_{i}^{2}}\;\int_{|\omega|\ll\tilde{\epsilon}}\!\!\!\!\!\! d\omega\;\tilde{G}_{\Gamma_{i}}(\omega)\;.\end{eqnarray}

We thus arrive at a correlation decay, for the Majorana modes, of
the form \begin{eqnarray}
\overline{\left\langle \gamma\left(0\right)\gamma\left(\mathrm{T}\right)\right\rangle } & \approx & \exp\left[-4\sum_{i}\frac{1}{\epsilon_{i}^{2}}\int_{|\omega|\ll\tilde{\epsilon}}\!\!\!\!\!\! d\omega\;\tilde{G}_{\Gamma_{i}}(\omega)\;\right]\;.\label{eq:rand-amplitude-low}\end{eqnarray}
 The coefficient in the exponent depends on the spectral weight of
the noise. From Parceval's theorem, $\int_{-\infty}^{\infty}d\omega\;\tilde{G}_{\Gamma_{i}}(\omega)=\overline{|\Gamma_{i}(t)|^{2}}$,
so the prefactor depends on the intensity of fluctuations of the couplings
$\Gamma_{i}(t)$ in time. When the fluctuations are large, for example
when the $\Gamma_{i}(t)$ are tied to thermally induced vibrations
in two dimensional systems, there is large decoherence.

We remark that even in the cases when $\sigma_{i}^{2}(\mathrm{T}\to\infty)$
is bounded, the value may be rather large, and the Majorana correlation
is \textit{exponential} in this value. Therefore keeping the error
to within reasonable bounds for quantum error correction to be applicable
can be a tall order. In this sense, the Majorana qubit is not necessarily
any more robust than other proposed qubit platforms.

\subsubsection*{Case II: High-frequency noise}

In this case we compute $\sigma_{i}^{2}(\mathrm{T})$ assuming that
the correlations $G_{\Gamma_{i}}(\tau_{1}-\tau_{2})$ decay in time,
so one can break the $\tau_{1,2}$ integrals into center of mass:
$(\tau_{1}+\tau_{2})/2$ and relative coordinates $\tau_{1}-\tau_{2}$
integrals, and in the limit of large $\mathrm{T}$ one has \begin{equation}
\sigma_{i}^{2}(\mathrm{T})\xrightarrow[{\mathrm{T\; large}}]{}2\mathrm{T}\;\tilde{G}_{\Gamma}(\epsilon_{i})\;,\label{eq:LongTimeCorrelatorMatrix}\end{equation}
 where $\tilde{G}_{\Gamma}(\epsilon_{i})$ is the Fourier transform
of $\mathrm{G_{\Gamma}(\tau)}$ at frequency $\epsilon_{i}$. We further
clarify this in Fig.~(\ref{fig:Correlator}).

We thus arrive at a correlation decay, for the Majorana modes, of
the form \begin{eqnarray}
\overline{\left\langle \gamma\left(0\right)\gamma\left(\mathrm{T}\right)\right\rangle } & \approx & \exp\left[-4\,\mathrm{T}\sum_{i}\tilde{G}_{\Gamma}(\epsilon_{i})\right]\;.\label{eq:rand-amplitude}\end{eqnarray}

Notice that this expression has meaning only if the $\tilde{G}_{\Gamma}(\omega)$
has spectral weight above the gap $\tilde{\epsilon}$. If not, one
has to treat the problem in the low frequency limit discussed above.

\subsubsection{\label{par:Non-ZeroValues}Non zero expectation values}

One can generalize this result for when the $\Gamma_{i}$ fluctuations
are centered around a non-zero value $\Gamma_{i}^{0}$. In this case,
\begin{equation}
P(\vec{Z}_{i})=\frac{1}{2\pi\sigma_{i}^{2}(\mathrm{T})}\;\exp\left(-\frac{1}{2}\frac{|\vec{Z}_{i}-\vec{Z}_{i}^{0}(\mathrm{T})|^{2}}{\sigma_{i}^{2}(\mathrm{T})}\right)\;\;,\label{eq:OffsetProbabilityDistribution}\end{equation}
 where \begin{equation}
Z_{i}^{0}(\mathrm{T})=\sqrt{2}\Gamma_{i}^{0}\;\int_{0}^{\mathrm{T}}d\tau\; e^{-i\epsilon_{i}\tau}=\sqrt{2}i\,\Gamma_{i}^{0}\;\frac{e^{-i\epsilon_{i}\mathrm{T}}-1}{\epsilon_{i}}\;\;,\label{eq:baredisplacementfield}\end{equation}
 which lead to \begin{eqnarray}
\mathrm{\overline{\left\langle \gamma\left(0\right)\gamma\left(\mathrm{T}\right)\right\rangle }} & = & \prod_{i}\frac{e^{-\frac{|Z_{i}^{0}(\mathrm{T})|^{2}}{1+2\sigma_{i}({\mathrm{T}})}}}{1+2\sigma_{i}({\mathrm{T}})}\;.\label{eq:DisplacementCoherence}\end{eqnarray}
 Notice that we recover the static result Eq.~(\ref{eq:non-int-static})
of the previous section if there is no disorder {[}$\sigma_{i}({\mathrm{T}})=0${]}.
Indeed we see that $\left\langle \gamma\left(0\right)\gamma\left(\mathrm{T}\right)\right\rangle =\prod_{i}e^{-|Z_{i}^{0}(\mathrm{T})|^{2}}$.

In the particular case of high-frequency noise (non-zero $G_{\Gamma_{i}}(\epsilon_{i})$),
one obtains in the large $\mathrm{T}$ limit one obtains\begin{equation}
\mathrm{\overline{\left\langle \gamma\left(0\right)\gamma\left(\mathrm{T}\right)\right\rangle }}\xrightarrow[{\mathrm{T\; large}}]{}\prod_{i}\left[1+4T\widetilde{G}_{\Gamma}\left(\epsilon_{i}\right)\right]^{-1},\label{eq:}\end{equation}
which agrees with the case where the fluctuations are centered around
zero shown in Eq.~(\ref{eq:rand-amplitude}).

\subsubsection{\label{sec:Cross-Correlations}Cross correlations of fluctuations}

We would now like to extend our model to include cross correlations
of fluctuations between the modes. Once again we focus on a Hamiltonian
of the form $H_{\mathrm{Mean}}=\gamma\sum_{i=1}^{N}\left(\Gamma_{i}c_{i}-\Gamma_{i}^{*}c_{i}^{\dagger}\right)+\sum_{i=1}^{N}\epsilon_{i}c_{i}^{\dagger}c_{i}$.
Here $\gamma$ is a single Majorana mode and $c_{i}$$\,$, $c_{i}^{\dagger}$
are regular fermion creation and annihilation operators. In our model
we will allow for Gaussian classical dynamics for the coupling constants
$\Gamma_{i}$ with possible cross correlations between the couplings.
More precisely, we will assume that the probability distribution of
couplings may be written as: \begin{widetext}

\begin{equation}
\mathrm{P}\left(\{\Gamma_{\mathrm{i}}(\tau),\Gamma_{\mathrm{i}}^{*}(\tau)\}\right)={\mathcal{Z}}^{-1}\;\int\int\mathcal{D}\left\{ \Gamma_{i}^{*}\left(\tau\right),\Gamma_{i}\left(\tau\right)\right\} \exp\left(\frac{-1}{2}\int_{-\infty}^{\infty}\int_{-\infty}^{\infty}d\tau_{1}d\tau_{2}\sum_{i,j}G_{i,j}^{-1}\left(\tau_{1},\tau_{2}\right)\Gamma_{i}^{*}\left(\tau_{1}\right)\Gamma_{j}\left(\tau_{2}\right)\right)\label{eq:CrossCorrelators}\end{equation}

Next we introduce the $\vec{\mathcal{Z}}\equiv\left(\mathcal{Z}_{1},.....\mathcal{Z}_{N}\right)\in\mathbb{C}^{N}$
with $\mathcal{Z}_{i}\left(\mathrm{T}\right)=\sqrt{2}\int_{0}^{\mathrm{T}}d\tau\;\Gamma_{i}(\tau)\;\mathrm{e}^{-\mathrm{i}\int_{0}^{\tau}dt\,\epsilon_{i}\left(t\right)}$.
With this notation we may write that:\begin{equation}
\left\langle \gamma\left(0\right)\gamma\left(T\right)\right\rangle =e^{-\vec{\mathcal{Z}}^{\dagger}\vec{\mathcal{Z}}}\label{eq:CoherenceNotation.}\end{equation}

Which is just a rewriting of Eq. (\ref{eq:ExpectationValueDisplacement}).
Next following Eq. (\ref{eq:CharacteristicFunction}) we may write
that:\textbf{\begin{eqnarray}
\tilde{P}\left(\vec{\mathcal{K}}\right) & = & \int d^{2}\mathcal{Z}_{1}\int d^{2}\mathcal{Z}_{2}\int d^{2}\mathcal{Z}_{3}....\int d^{2}\mathcal{Z}_{N}\; P(\vec{\mathcal{Z}})\; e^{\frac{-i}{2}\left(\vec{\mathcal{Z}}^{\dagger}\vec{\mathcal{K}}+\vec{\mathcal{K}}^{\dagger}\vec{\mathcal{Z}}\right)}\nonumber \\
 & = & {\mathcal{N}}^{-1}\int\mathcal{D}\Gamma_{i}\,\mathcal{D}\Gamma_{i}^{*}\;\;\mathrm{e}^{-\frac{1}{2}\int_{-\infty}^{\infty}d\tau_{1}\int_{-\infty}^{\infty}d\tau_{2}\;\Gamma_{i}^{*}(\tau_{1})\; G_{ij}^{-1}(\tau_{1},\tau_{2})\;\Gamma_{j}(\tau_{2})}\nonumber \\
 &  & \qquad\qquad\qquad\times e^{-i\frac{1}{2}\,\sqrt{2}\sum_{i}{\mathcal{K}}_{i}^{*}\int_{0}^{\mathrm{T}}d\tau\;\Gamma_{i}(\tau)\; e^{-i\epsilon_{i}\tau}}\;\times e^{-i\frac{1}{2}\,\sqrt{2}\sum_{i}{\mathcal{K}}_{i}\int_{0}^{\mathrm{T}}\int_{0}^{\mathrm{T}}d\tau\;\Gamma_{i}^{*}(\tau)\; e^{+i\epsilon_{i}\tau}}\nonumber \\
 & = & \exp\left(-\frac{1}{2}\,\times2\times\sum_{i,j}{\mathcal{K}}_{i}^{*}{\mathcal{K}}_{j}\int_{0}^{\mathrm{T}}\!\!\!\! d\tau_{1}\;\int_{0}^{\mathrm{T}}\!\!\!\! d\tau_{2}\;\; e^{-i\epsilon_{i}\tau_{1}}\; G_{ij}(\tau_{1},\tau_{2})\; e^{+i\epsilon_{j}\tau_{2}}\right)\;.\label{eq:CharacteristicFunctionVector}\end{eqnarray}
}  \end{widetext}

From this equation we see that the distribution $P\left(\vec{\mathcal{Z}}\right)$
is a Gaussian with a covariance matrix $\boldsymbol{\sigma}\left(\mathrm{T}\right)$
given by:\begin{equation}
\boldsymbol{\sigma_{ij}}\left(\mathrm{T}\right)\equiv2\int_{0}^{\mathrm{T}}\!\!\!\! d\tau_{1}\;\int_{0}^{\mathrm{T}}\!\!\!\! d\tau_{2}\;\; e^{-i\epsilon_{i}\tau_{1}}\; G_{ij}(\tau_{1},\tau_{2})\; e^{+i\epsilon_{j}\tau_{2}}\label{eq:Covariance matrix}\end{equation}

Combining and simplifying we may write that:\begin{equation}
\left\langle \gamma\left(0\right)\gamma\left(\mathrm{T}\right)\right\rangle =\frac{1}{\det\left(\mathbb{I}+2\boldsymbol{\sigma}\left(\mathrm{T}\right)\right)}\label{eq:FinalValueCorrelated}\end{equation}

Here $\mathbb{I}$ is the identity matrix ($\mathbb{I}_{ij}=\delta_{ij}$).
We can also generalize to the case where the couplings have a non-zero
expectation value, $\Gamma_{i}=\Gamma_{i}^{0}+\delta\Gamma_{i}$,
with the $\delta\Gamma_{i}$ having a probability distribution given
by Eq. (\ref{eq:CrossCorrelators}). In this case, we obtain:\begin{equation}
\left\langle \gamma\left(0\right)\gamma\left(\mathrm{T}\right)\right\rangle =\frac{\exp\left(-{\vec{\mathcal{Z}}}_{0}^{\dagger}\left(\mathrm{T}\right)\left(\mathbb{I}+2\boldsymbol{\sigma}\left(\mathrm{T}\right)\right)^{-1}\vec{\mathcal{Z}}_{0}\left(\mathrm{T}\right)\right)}{\det\left(\mathbb{I}+2\boldsymbol{\sigma}\left(\mathrm{T}\right)\right)}\label{eq:OffsetCrossCorrelated}\end{equation}

Here, similarly to Section~\ref{par:Non-ZeroValues}, we have introduced
the vector $\vec{\mathcal{Z}}_{0}$ whose i'th component is given
by: $\mathcal{Z}_{0,i}\left(\mathrm{T}\right)=\sqrt{2}i\,\Gamma_{i}^{0}\;\frac{e^{-i\epsilon_{i}\mathrm{T}}-1}{\epsilon_{i}}$.

\subsection{\label{sub:Fluctuating-Energies}Fluctuating energies}

Let us consider the case where the energies undergo Gaussian fluctuations
in time, around some average value: $\mathrm{\epsilon_{i}(\tau)=\epsilon_{i}+\delta\epsilon_{i}(\tau)}$
with $\left\langle \delta\epsilon_{i}\left(\tau_{1}\right)\delta\epsilon_{i}\left(\tau_{2}\right)\right\rangle =G_{i}\left(\tau_{1},\:\tau_{2}\right)$.
Let $\varphi(\tau)\equiv\int_{0}^{\tau}dt\;\delta\epsilon_{i}(t)$.
If the $\delta\epsilon_{i}(\tau)$ are short-time correlated the quantity
$\mathrm{\overline{\left[\varphi(\tau_{1})-\varphi(\tau_{2})\right]^{2}}\equiv G_{\varphi}^{2}(\tau_{1}-\tau_{2})}$
will grow linearly in $|\tau_{1}-\tau_{2}|$. We note that the phases
$\varphi_{i}\left(\tau\right)$ execute random walks in this case.

The magnitude square of the {}``position'' of the $Z_{i}$ has average
\begin{widetext}\begin{eqnarray}
\overline{|Z_{i}(\mathrm{T})|^{2}} & =2 & |\Gamma_{i}|^{2}\;\int_{0}^{\mathrm{T}}\!\!\!\! d\tau_{+}\;\int_{0}^{\mathrm{T}}\!\!\!\! d\tau_{-}\;\; e^{+i\epsilon_{i}\tau_{+}}\;\overline{e^{+i[\varphi(\tau_{+})-\varphi(\tau_{-})]}}\; e^{-i\epsilon_{i}\tau_{-}}\nonumber \\
 & =2 & |\Gamma_{i}|^{2}\;\int_{0}^{\mathrm{T}}\!\!\!\! d\tau_{+}\;\int_{0}^{\mathrm{T}}\!\!\!\! d\tau_{-}\;\; e^{+i\epsilon_{i}\tau_{+}}\; e^{-\frac{1}{2}G_{\varphi}^{2}(\tau_{+}-\tau_{-})}\; e^{-i\epsilon_{i}\tau_{-}}\;.\label{eq:TwoPointEnergy}\end{eqnarray}
 The calculation of higher moments is quite similar if the term $\mathrm{e^{G_{\varphi}(\tau_{+}-\tau_{-})}}$
confines the two times to be close to each other.

\begin{eqnarray}
\overline{\left|Z_{i}\left(\mathrm{T}\right)\right|^{2n}} & = & 2^{n}|\Gamma_{i}|^{2n}\;\int_{0}^{\mathrm{T}}\!\!\!\!\mathrm{d\tau_{1}^{+}\;\dots\int_{0}^{\mathrm{T}}\!\!\!\! d\tau_{n}^{+}\;\int_{0}^{\mathrm{T}}\!\!\!\! d\tau_{1}^{-}\;\dots\int_{0}^{\mathrm{T}}\!\!\!\! d\tau_{n}^{-}\;\;}e^{i\epsilon_{i}\sum_{j}\tau_{j}^{+}}\:\overline{e^{i\sum_{j}\varphi(\tau_{j}^{+})-i\sum_{j}\varphi(\tau_{j}^{-})]}}\;\;\mathrm{e}^{-i\epsilon_{i}\sum_{j}\tau_{j}^{-}}\;\nonumber \\
 & = & 2^{n}|\Gamma_{i}|^{2n}\;\int_{0}^{\mathrm{T}}\!\!\!\!\mathrm{d\tau_{1}^{+}\;\dots\int_{0}^{\mathrm{T}}\!\!\!\! d\tau_{n}^{+}\;\int_{0}^{\mathrm{T}}\!\!\!\! d\tau_{1}^{-}\;\dots\int_{0}^{\mathrm{T}}\!\!\!\! d\tau_{n}^{-}\;\;}e^{i\epsilon_{i}\sum_{j}\tau_{j}^{+}}\times\mathrm{e}^{-i\epsilon_{i}\sum_{j}\tau_{j}^{-}}\nonumber \\
 & \times & \exp\left[-\frac{1}{2}\int_{0}^{\tau_{1}^{+}}du_{1}..\int_{0}^{\tau_{n}^{+}}du_{n}\int_{0}^{\tau_{1}^{-}}dv_{1}..\int_{0}^{\tau_{n}^{-}}dv_{n}\:\sum_{i=1}^{n}\left\{ G\left(u_{i},\, u_{j}\right)+G\left(v_{i},\, v_{j}\right)-G\left(u_{i},\, v_{j}\right)-G\left(v_{i},\, u_{j}\right)\right\} \right]\nonumber \\
 & \cong & 2^{n}\left|\Gamma_{i}\right|^{2n}\: n!\,\left(\int_{0}^{\mathrm{T}}\!\!\!\! d\tau_{+}\;\int_{0}^{\mathrm{T}}\!\!\!\! d\tau_{-}\;\; e^{+i\epsilon_{i}\tau_{+}}\; e^{-\frac{1}{2}G_{\varphi}^{2}(\tau_{+}-\tau_{-})}\; e^{-i\epsilon_{i}\tau_{-}}\right)^{n}\nonumber \\
 & = & n!\:\left(\overline{\left|Z_{i}\left(\mathrm{T}\right)\right|^{2}}\right)^{n}\:.\label{eq:HighPowerCorrelator}\end{eqnarray}

For the second equality we have used the fact that the process is
Gaussian. In this way we mapped the problem to the partition function
of a two species Coulomb like gas. Then in the fourth line we have
used a dipole approximation for the partition function. We note that
this is consistent with the confining assumption as $\int_{\tau_{1}}^{\tau_{2}}\int_{\tau_{1}}^{\tau_{2}}dudvG\left(u,v\right)\propto\left|\tau_{1}-\tau_{2}\right|$
so that we have a confining linear potential between oppositely charged
particles of our Coulomb gas.

We now claim that $Z_{i}$ will execute diffusion because of the random
phases. Indeed, these correlation functions are the moments of a Gaussian
distribution with variance $\overline{\left|Z_{i}\left(\mathrm{T}\right)\right|^{2}}$.
This variance can often be computed in the high-frequency case (similarly
to Section~\ref{sub:Fluctuating-amplitudes}) and for large $\mathrm{T}$
one can approximate\begin{equation}
\overline{\left|Z_{i}(\mathrm{T})\right|^{2}}\xrightarrow[{\mathrm{T\; large}}]{}2\mathrm{T}\;|\Gamma_{i}|^{2}\;\int_{-\infty}^{\infty}\!\!\!\! d\tau\;\; e^{+i\epsilon_{i}\tau}\; e^{-\frac{1}{2}G_{\varphi}^{2}(\tau)}\;\;\equiv\mathrm{T}\Theta_{i}\;,\label{eq:AverageFluctuation}\end{equation}
 and the probability distribution is given by $\mathrm{P}\left(Z_{i}\left(\mathrm{T}\right)\right)\cong\frac{1}{2\pi\Theta_{i}^{2}(\mathrm{T})}\;\exp\left(-\frac{1}{2}\frac{|Z_{i}\left(\mathrm{T}\right))|^{2}}{\mathrm{T}\Theta_{i}^{2}(\mathrm{T})}\right)$.
Repeating the analysis of Section~\ref{sub:Fluctuating-amplitudes},
we get a power law decay (for each mode $i$) for the coherence of
Majorana qubit, with a coefficient that is dependent on the Fourier
transform of the exponential of the $G_{\varphi}^{2}(\tau)$ correlation
function: \begin{eqnarray}
\overline{\left\langle \gamma\left(0\right)\gamma\left(\mathrm{T}\right)\right\rangle } & = & \prod_{\mathrm{i}}\left[1+4\,\mathrm{T}\;|\Gamma_{i}|^{2}\;\int_{-\infty}^{\infty}\!\!\!\! d\tau\;\; e^{+i\epsilon_{i}\tau}\; e^{-\frac{1}{2}G_{\varphi}^{2}(\tau)}\right]^{-1}\nonumber \\
 & \approx & \exp\left[-4\,\mathrm{T}\sum_{i}|\Gamma_{i}|^{2}\;\int_{-\infty}^{\infty}\!\!\!\! d\tau\;\; e^{+i\epsilon_{i}\tau}\; e^{-\frac{1}{2}G_{\varphi}^{2}(\tau)}\right]\;.\label{eq:rand-energy}\end{eqnarray}

\end{widetext}

For $G_{\varphi}^{2}(\tau)\propto|\tau|$, the Fourier transform of
$e^{-\frac{1}{2}G_{\varphi}^{2}(\tau)}$ will decay as a power law
in frequency. We would like to point out that if the $\mathrm{\epsilon_{i}(\tau)}$
have a correlation time $\tau_{\Omega}=\Omega^{-1}$, the short-time
behavior of $G_{\varphi}^{2}(\tau)$ is smoothened, and the kink-singularity
of at $\tau=0$ disappears, while the long-time behavior $|\tau|$
remains the same. Using general results on Fourier transforms \cite{key-55}
we know that the Fourier transform of $e^{-\frac{1}{2}G_{\varphi}^{2}(\tau)}$
will decay faster than any power of frequency $\omega$ when $\omega\gg\Omega$.
This indicates a good level of protection for systems with large gaps
compared to the bandwidth of the noise source.

\subsection{\label{sub:Telegraph-Noise}Telegraph noise fluctuations of coupling
amplitudes}

Here we shall study classical telegraphic noise. Our model for telegraphic
noise will be a $\Gamma_{i}(\tau)$ that switches between $\pm\Lambda_{i}$
with time intervals between events that are distributed randomly with
characteristic frequency $\Omega_{i}^{-1}$. The complex number $Z_{i}(\mathrm{T})$
will again perform a random walk at long times, which we will confirm
by computing the moments of $|Z_{i}(\mathrm{T})|^{2}$. Let us start
by computing the second moment: \begin{widetext} \begin{eqnarray}
\overline{\mathrm{|Z_{i}(\mathrm{T})|^{2}}} & =2 & \int_{0}^{\mathrm{T}}\!\!\!\! d\tau_{+}\;\int_{0}^{\mathrm{T}}\!\!\!\! d\tau_{-}\;\; e^{+i\epsilon_{i}\tau_{+}}\;\overline{\Gamma_{i}(\tau_{+})\;\Gamma_{i}(\tau_{-})}\; e^{-i\epsilon_{i}\tau_{-}}\;.\label{eq:TelegraphNoiseTwoPoint}\end{eqnarray}
 Now, $\mathrm{|Z_{i}(\mathrm{T})|^{2}}=2\Lambda_{i}^{2}\;(-1)^{N_{\mathrm{flips}}(\tau_{-},\tau_{+})}$,
where $N_{\mathrm{flips}}(\tau_{-},\tau_{+})$ is the number of switches
between the two times $\tau_{\pm}$. The average \begin{eqnarray}
\overline{(-1)^{N_{\mathrm{flips}}(\tau_{-},\tau_{+})}} & = & \sum_{N=0}^{\infty}(-1)^{N}\;\frac{1}{N!}\left(\Omega_{i}\,|\tau_{+}-\tau_{-}|\right)^{N}\; e^{-\Omega_{i}\,|\tau_{+}-\tau_{-}|}\nonumber \\
 & = & \; e^{-2\,\Omega_{i}\,|\tau_{+}-\tau_{-}|}\;,\label{eq:TelegraphNoiseflips}\end{eqnarray}
 so we obtain\begin{equation}
\overline{|\mathrm{Z_{i}(\mathrm{T})}|^{2}}\xrightarrow[{\mathrm{T\; large}}]{}2\mathrm{\mathrm{T}}\;\Lambda_{i}^{2}\;\frac{4\,\Omega_{i}}{(2\,\Omega_{i})^{2}+\epsilon_{i}^{2}}\;.\end{equation}

In the appendix we compute the higher moments and show that the distribution
of $Z_{i}(\mathrm{T})$ approaches a Gaussian, as intuitively expected
from the fact that the telegraph noise causes the fictitious particle
position to diffuse at times larger compared to the switching time.
We obtain, similarly to the previous cases discussed above, that\begin{eqnarray}
\overline{\left\langle \gamma\left(0\right)\gamma\left(\mathrm{T}\right)\right\rangle } & = & \prod_{\mathrm{i}}\left[1+2\mathrm{T}\;\Lambda_{i}^{2}\;\frac{4\,\Omega_{i}}{(2\,\Omega_{i})^{2}+\epsilon_{i}^{2}}\right]^{-1}\nonumber \\
 & \approx & \exp\left[-2\mathrm{T}\;\sum_{i}\;\Lambda_{i}^{2}\;\frac{4\,\Omega_{i}}{(2\,\Omega_{i})^{2}+\epsilon_{i}^{2}}\right]\;.\label{eq:telegraph-noise}\end{eqnarray}

\end{widetext}In the last line we assumed that there are many relevant
fluctuating levels each making a small contribution so that we are
able to re-exponentiate. From this we see that due to the effects
of telegraph noise the information stored in the Majorana qubit is
lost on a time scale $\sim{\tau_{\mathrm{typ}}}/{\sum_{i}\frac{\left|\Lambda_{i}\right|^{2}}{\epsilon_{i}^{2}}}$.
Here $\mathrm{\tau_{typ}}\sim\Omega^{-1}$ is the typical switching
rate for the regular fermion modes. This is an exponential decay of
Majorana coherence with the rate given by a rational function of the
the coupling strengths and frequencies of the switching. This leads
to short lifetimes of Majorana modes. We would like to note that the
power law term comes from the instantaneous switching process. For
a finite switching speed and as such a smooth $\left\langle \Gamma\left(\tau\right)\Gamma\left(v\right)\right\rangle $
the Fourier transform in Eq. (\ref{eq:telegraph-noise}) would decay
faster then any rational function of $\epsilon_{i}$ for large $\epsilon_{i}$
(as compared to the inverse switching time)\cite{key-55}.

\section{Conclusions\label{sec:Conclusions}}

In this work we have studied the stability of qubits constructed from
Majorana zero modes, for example using an encoding such as $\sigma^{z}=i\gamma_{1}\gamma_{2}$.
The persistence of memory can be measured from two-time correlations
such as $\left\langle \sigma^{z}\left(0\right)\sigma^{z}\left(\mathrm{T}\right)\right\rangle $,
which we have shown is independent of the particular state of the
qubit. We have shown that the if the environments coupling to each
Majorana mode are uncorrelated, then the qubit overlap function factorizes:
$\left\langle \sigma^{z}\left(0\right)\sigma^{z}\left(\mathrm{T}\right)\right\rangle =\left\langle \gamma_{1}\left(0\right)\gamma_{1}\left(\mathrm{T}\right)\right\rangle \left\langle \gamma_{2}\left(0\right)\gamma_{2}\left(\mathrm{T}\right)\right\rangle $.
We then analyzed, in detail, the decay of the Majorana two-point function
$\left\langle \gamma\left(0\right)\gamma\left(\mathrm{T}\right)\right\rangle $,
when the Majoranas couple via tunneling to fermions in a bath. We
considered only baths where the fermions had a gapped single particle
spectrum (gapless baths would trivially destroy coherence). We considered
both cases where the tunneling amplitudes were static, and cases where
they were dynamical, fluctuating either classically or quantum mechanically,
say mediated by a boson bath.

Static tunnelings are, expectantly, not consequential leading to finite
decay. Though this serves as a way to check our generic formalism.
More precisely if the fermions in the bath are non-interacting and
if the tunnelings are just switched on but then kept constant thereafter,
then the Majorana qubits only experience a finite depletion which
we checked by explicitly rediagonalizing the non-interacting fermionic
Hamiltonian with the new couplings. This result can be easily interpreted
as a finite adjustment in the overlap of the qubit before and after
the basis changes upon switching the tunnelings.

However, dynamic fluctuations of the tunneling amplitudes can have
very serious consequences. Our analysis makes it clear that the dephasing
of the Majorana correlations is tied hand-in-hand to fluctuations
(spectral functions) of both the fermionic bath and the noise. In
some instances, for example in the case of athermal telegraphic noise,
fluctuations can destroy the Majorana memories, leading to complete
decay of coherence at long times. We analyzed several types of noise
in the bath, both classical and quantum. To understand the rate of
information loss in experimentally relevant systems it is important
to study various materials, relevant sources of noise and in general
realistic spectral functions of the bath. The formalism here presented
forms the basis for such analysis.

\section*{Acknowledgments}

We gratefully acknowledge useful discussions with Bert Halperin, Chang-Yu
Hou, Chris Laumann, Dung-Hai Lee, Patrick Lee, Eduardo Mucciolo, Christopher
Mudry, Andrew Potter, Shinsey Ryu, Jay Deep Sau, Michael Stone, and
Xiao-Gang Wen. This work was supported by NSF Grant CCF-1116590.

\vspace{0.5cm}

\appendix

\section{Non interacting systems (quantum depletion)\label{sec:QuadraticHamiltonian}}

To have yet another independent check of the results presented in
the paper we would like to derive results similar to Eq. (\ref{eq:LongTimeStatic})
in a different way. More precisely we will consider a model consisting
of a Majorana mode interacting via tunneling with non-interacting
complex fermionic modes. The Hamiltonian of our system will be:\begin{equation}
H_{\mathrm{Mean}}=\gamma\sum_{i=1}^{N}\left(\Gamma_{i}c_{i}-\Gamma_{i}^{*}c_{i}^{\dagger}\right)+\sum_{i=1}^{N}\epsilon_{i}c_{i}^{\dagger}c_{i}\label{eq:HamiltonianQuadraticTunneling}\end{equation}

We will first proceed by exactly re-diagonalizing the Hamiltonian.
By taking commutators of the form $\left[H_{\mathrm{Mean}},\,\gamma\right]$,
$\left[H_{\mathrm{Mean}},\, c_{i}\right]$ and $\left[H_{\mathrm{Mean}},\, c_{i}^{\dagger}\right]$
we may rewrite this Hamiltonian as a matrix acting on the space spanned
by $\left\{ \frac{\gamma}{\sqrt{2}},\, c_{i},\, c_{i}^{\dagger}\right\} $
(the factor of $\sqrt{2}$ is a normalization constant that insures
that the matrix representing the Hamiltonian is Hermitian in this
basis). With respect to this basis we may write that:\begin{widetext}\begin{equation}
H_{\mathrm{Mean}}=\mathrm{\left(\begin{array}{ccccccccc}
0 & \sqrt{2}\Gamma_{1} & \cdots & \cdots & \sqrt{2}\Gamma_{N} & -\sqrt{2}\Gamma_{1}^{*} & \cdots & \cdots & -\sqrt{2}\Gamma_{N}^{*}\\
\sqrt{2}\Gamma_{1}^{*} & \epsilon_{1} & 0 & \cdots & 0 & 0 & \cdots & \cdots & 0\\
\vdots & 0 & \epsilon_{2} & \ddots & \vdots & \vdots &  &  & \vdots\\
\vdots & \vdots & \ddots & \ddots & 0 & \vdots &  &  & \vdots\\
\sqrt{2}\Gamma_{N}^{*} & 0 & \cdots & 0 & \epsilon_{N} & 0 & \cdots & \cdots & 0\\
-\sqrt{2}\Gamma_{1} & 0 & \cdots & \cdots & 0 & -\epsilon_{1} & 0 & \cdots & 0\\
\vdots & \vdots &  &  & \vdots & 0 & -\epsilon_{2} & \ddots & \vdots\\
\vdots & \vdots &  &  & \vdots & \vdots & \ddots & \ddots & 0\\
-\sqrt{2}\Gamma_{N} & 0 & \cdots & \cdots & 0 & 0 & \cdots & 0 & -\epsilon_{N}\end{array}\right)}\label{eq:MatrixHamiltonian}\end{equation}
 \end{widetext} We may now diagonalize this matrix by solving for
the eigenvalues of the system $\left\{ \lambda_{\kappa}\right\} $
with corresponding eigenvectors $\left\{ V_{\kappa}\equiv U_{\kappa}\gamma+\sum_{i=1}^{N}U_{\kappa,i}c_{i}+\sum_{i=1}^{N}U_{\kappa,N+i}c_{i}^{\dagger}\right\} $.
By direct substitution into the equation $HV_{\kappa}=\lambda_{\kappa}V_{\kappa}$
we see that:\begin{eqnarray}
U_{\kappa,i} & = & \frac{\sqrt{2}\Gamma_{i}^{*}}{\lambda_{\kappa}-\epsilon_{i}}U_{\kappa},\label{eq:FirstMatrixEquations}\\
U_{\kappa,N+i} & = & -\frac{\sqrt{2}\Gamma_{i}}{\lambda_{\kappa}+\epsilon_{i}}U_{\kappa}\nonumber \end{eqnarray}

Here we have ignored the {}``top line'' of $H_{\mathrm{Mean}}$
in Eq. (\ref{eq:MatrixHamiltonian}). Substituting Eq. (\ref{eq:FirstMatrixEquations})
into the {}``top line'' of $H_{\mathrm{Mean}}$ we get that:\begin{eqnarray}
\lambda_{\kappa}U_{\kappa} & = & \sum_{i}\sqrt{2}\Gamma_{i}U_{\kappa,i}-\sum_{i}\sqrt{2}\Gamma_{i}^{*}U_{\kappa,N+i}\label{eq:IntermediateStepSolutionMatrixEquation}\\
 & = & \sum_{i}\frac{4\lambda_{\kappa}\left|\Gamma_{i}\right|^{2}}{\left(\lambda_{\kappa}\right)^{2}-\left(\epsilon_{i}\right)^{2}}U_{\kappa},\label{eq:SolutionMatrixEquations}\end{eqnarray}

We can now obtain eigenvalue equations:

\begin{equation}
\lambda_{\kappa}=0,\;\mathrm{or}\;1=4\sum_{i}\frac{\left|\Gamma_{i}\right|^{2}}{\left(\lambda_{\kappa}\right)^{2}-\left(\epsilon_{i}\right)^{2}}\label{eq:EigenValueEquations}\end{equation}

Now substituting $\lambda_{0}=0$ into Eq. (\ref{eq:IntermediateStepSolutionMatrixEquation})
we get that:\begin{eqnarray}
1 & = & \left|U_{0}\right|^{2}+\sum_{i=1}^{N}\left|U_{0,i}\right|^{2}+\sum_{i=1}^{N}\left|U_{0,N+i}\right|^{2}\nonumber \\
 & = & \left|U_{0}\right|^{2}\left(1+4\sum_{i=1}^{N}\frac{\left|\Gamma_{i}\right|^{2}}{\epsilon_{i}^{2}}\right)\label{eq:SumBounds}\end{eqnarray}

From this we see that the overlap of the new zero mode with the original
mode stays finite (which would lead to non-zero coherence for arbitrarily
long times) whenever:\begin{equation}
\sum_{i=1}^{N}\frac{\left|\Gamma_{i}\right|^{2}}{\epsilon_{i}^{2}}<\infty\label{eq:MainCondition}\end{equation}
 This result is similar to Eq. (\ref{eq:LongTimeStatic}) in the main
text. This condition is true for any finite system. However the overlap
of this mode with the original zero energy mode is depleted by a factor
of:\begin{eqnarray}
\left(1+\sum_{i=1}^{N}\left|U_{0,i}\right|^{2}+\sum_{i=1}^{N}\left|U_{0,N+i}\right|^{2}\right)^{-1/2}\nonumber \\
=\left(1+4\sum_{i}\frac{\left|\Gamma_{i}\right|^{2}}{\epsilon_{i}^{2}}\right)^{-1/2}.\label{eq:Deplition}\end{eqnarray}
 Below in Appendix \ref{sub:Summation-of-Eq.SumBounds} we will show
that this will remain so for mean field like infinite systems.

\section{\label{sub:Quantum-Fluctuations}Quantum fluctuations}

We would like to extend the previous results, see Section \ref{sec:Fluctuating-Hamiltonians},
to the case where the couplings $\Gamma_{i}$ are allowed to have
quantum fluctuations. That is we will allow for different fluctuations
for the backwards and forwards time paths. Once again we will focus
on a single Majorana mode which may be well described by a Hamiltonian
of the form $H_{\mathrm{Mean}}\left(\Gamma_{i},\:\Gamma_{i}^{*}\right)=\gamma\sum_{i=1}^{N}\left(\Gamma_{i}c_{i}-\Gamma_{i}^{*}c_{i}^{\dagger}\right)+\sum_{i=1}^{N}\epsilon_{i}c_{i}^{\dagger}c_{i}$.
Here $\gamma$ is a single Majorana mode and $c_{i}$, $c_{i}^{\dagger}$
are regular fermion creation and annihilation operators. In our model
we will allow for Gaussian quantum dynamics for the coupling constants
$\Gamma_{i}$. We will not be able to emulate the diffusion equation
derivation given in Section \ref{sub:Interacting-baths-and} but we
will provide a brute force resummation of the leading order terms
contributing to coherence. The key difficulty in modifying the approach
of Section\ref{sub:Interacting-baths-and} to the case of quantum
noise is that because of the various theta functions, see e.g. Eqs.
(\ref{eq:TimeOrderedEquations}) \& (\ref{eq:FinalValue}), the fermionic
part of the correlation function cannot be written in a factorisable
form $G_{F}\left(\tau_{1},\:\tau_{2}\right)\neq\widetilde{G_{1F}}\left(\tau_{1}\right)\times\widetilde{G_{2F}}\left(\tau_{2}\right)$
(or a sum of such terms). As such we cannot simply study the diffusion
of one or several modes, see e.g. Eq. (\ref{eq:Displacement}), but
we have to study the diffusion of an infinite number of degrees of
freedom (which is more difficult). We now proceed with the computation,
by using Eq. (\ref{eq:TimeOrderedEquations}) we may write that:\begin{widetext}\begin{equation}
\begin{array}{ccl}
\left\langle \gamma\left(0\right)\gamma\left(T\right)\right\rangle  & = & \mathcal{N}\int\int\mathcal{D}\left\{ \mathbf{\Gamma}^{\dagger},\mathbf{\Gamma}\right\} \exp\left(-\frac{1}{2}\,{\sum_{a,b}\;\int_{0}^{\mathrm{T}}\!\! d\tau_{1}^{a}\int_{0}^{\mathrm{T}}\!\! d\tau_{2}^{b}\;\;\mathbf{\Gamma}^{\dagger}\;\left(G_{F}^{(2)}\left(\tau_{1}^{a},\tau_{2}^{b}\right)\right)^{-1}\;\mathbf{\Gamma}}\right)\times\\
 &  & \times\gamma\exp\left(i\widetilde{\mathcal{T}}\int_{0}^{\mathrm{T}}\left\{ H_{\mathrm{Mean}}\left(\mathbf{\Gamma}^{\dagger}\left(\tau\right),\mathbf{\Gamma}\left(\tau\right)\right)\right\} d\tau\right)\gamma\exp\left(-i\mathcal{T}\int_{0}^{\mathrm{T}}\left\{ H_{\mathrm{Mean}}\left(\mathbf{\Gamma}^{\dagger}\left(\tau\right),\mathbf{\Gamma}\left(\tau\right)\right)\right\} d\tau_{2}\right)\\
 & = & \mathcal{N}\int\int\mathcal{D}\left\{ \mathbf{\Gamma}^{\dagger},\mathbf{\Gamma}\right\} \exp\left(-\frac{1}{2}\,{\sum_{a,b}\;\int_{0}^{\mathrm{T}}\!\! d\tau_{1}^{a}\int_{0}^{\mathrm{T}}\!\! d\tau_{2}^{b}\;\;\mathbf{\Gamma}^{\dagger}\;\left(\overline{G}_{F}^{(2)}\left(\tau_{1}^{a},\tau_{2}^{b}\right)\right)^{-1}\;\mathbf{\Gamma}}\right)\times\\
 &  & \times\exp\left(-\frac{1}{2}\,{\sum_{a,b}\;\int_{0}^{\mathrm{T}}\!\! d\tau_{1}^{a}\int_{0}^{\mathrm{T}}\!\! d\tau_{2}^{b}\;\;\mathbf{\Gamma}^{\dagger}\;\overline{D}_{\mathrm{F}}^{(2)}\left(\tau_{1}^{a},\tau_{2}^{b}\right)\;\mathbf{\Gamma}}\right)\end{array}\label{eq:MatrixKeldish}\end{equation}

Here $\mathrm{G}_{F}^{(2)}=\otimes_{i}\left(\begin{array}{cc}
G_{11}^{i}\left(\tau_{1},\tau_{2}\right) & G_{12}^{i}\left(\tau_{1},\tau_{2}\right)\\
G_{21}^{i}\left(\tau_{1},\tau_{2}\right) & G_{22}^{i}\left(\tau_{1},\tau_{2}\right)\end{array}\right)$, $\mathcal{N}=\det G_{F}^{\left(2\right)}$ and $\overline{D}_{\mathrm{F}}^{(2)}\left(\tau_{1}^{a},\tau_{2}^{b}\right)$
was defined in Eq. (\ref{eq:FinalKeldish}). We note that Eq. (\ref{eq:master-corgeneralr})
does not apply as there are correlations between the $\Gamma$'s.
As such we must compute a functional determinant as shown in Eq. (\ref{eq:MatrixKeldish})
above. We now use the equation:\begin{equation}
\int\int dz_{1}...dz_{n}dz_{1}^{*}...dz_{n}^{*}\exp\left(-\frac{1}{2}\vec{z}^{\dagger}G^{-1}\vec{z}\right)=\left(2\pi\right)^{n}\det\left(\mathrm{G}\right)\label{eq:Determinant}\end{equation}
 Which is true even for an arbitrary (not necessarily Hermitian) matrix
$G$. We will provide an independent proof of this result in Appendix
\ref{sec:Tedious-Calculations}. Now noting that the determinant of
a block diagonal matrix factorizes and writing out the form of $\overline{D}_{\mathrm{F}}^{(2)}\left(\tau_{1}^{a},\tau_{2}^{b}\right)$
say by using Eq. (\ref{eq:TimeOrderedEquations}) we can show that:\begin{equation}
\left\langle \gamma\left(0\right)\gamma\left(\mathrm{T}\right)\right\rangle ^{-1}=\prod_{\mathrm{i}}\det\left\{ \mathbb{I}+2\left(\begin{array}{cc}
G_{11}^{i}\left(\tau_{1},\tau_{2}\right) & G_{12}^{i}\left(\tau_{1},\tau_{2}\right)\\
G_{21}^{i}\left(\tau_{1},\tau_{2}\right) & G_{22}^{i}\left(\tau_{1},\tau_{2}\right)\end{array}\right)\left(\begin{array}{cc}
{\theta\left(t_{1}-t_{2}\right)\left\langle c_{i}^{\dagger}\left(t_{1}\right)c_{i}\left(t_{2}\right)\right\rangle \atop +\theta\left(t_{2}-t_{1}\right)\left\langle c_{i}\left(t_{2}\right)c_{i}^{\dagger}\left(t_{1}\right)\right\rangle } & \left\langle c_{i}^{\dagger}\left(t_{1}\right)c_{i}\left(t_{2}\right)\right\rangle \\
\left\langle c_{i}\left(t_{2}\right)c_{i}^{\dagger}\left(t_{1}\right)\right\rangle  & {\theta\left(t_{2}-t_{1}\right)\left\langle c_{i}^{\dagger}\left(t_{1}\right)c_{i}\left(t_{2}\right)\right\rangle \atop +\theta\left(t_{1}-t_{2}\right)\left\langle c_{i}\left(t_{2}\right)c_{i}^{\dagger}\left(t_{1}\right)\right\rangle }\end{array}\right)\right\} \label{eq:FinalValue}\end{equation}

We have inserted the forms of the various matrices explicitly. What
remains is to evaluate the functional determinant in Eq. (\ref{eq:FinalValue})
above. First by conjugating all matrices above with the matrix $\frac{1}{\sqrt{2}}\left(\begin{array}{cc}
\mathbb{I} & \mathbb{I}\\
\mathbb{I} & -\mathbb{I}\end{array}\right)$ (here $\mathbb{I}$ stands for the identity matrix on $\left[0,\mathrm{T}\right]\times\left[0,\mathrm{T}\right]$)
we may write that:\begin{equation}
\begin{array}{l}
\left\langle \gamma\left(0\right)\gamma\left(\mathrm{T}\right)\right\rangle ^{-1}=\\
\prod_{i}\det\left\{ \mathbb{I}+2\left(\begin{array}{cc}
0 & G_{i}^{R}\\
G_{i}^{A} & G_{i}^{K}\end{array}\right)\left(\begin{array}{cc}
0 & {\theta\left(t_{1}-t_{2}\right)\left\{ \left\langle c_{i}^{\dagger}\left(t_{1}\right)c_{i}\left(t_{2}\right)\right\rangle \right.\atop \qquad\qquad\left.-\left\langle c_{i}\left(t_{2}\right)c_{i}^{\dagger}\left(t_{1}\right)\right\rangle \right\} }\\
{\theta\left(t_{2}-t_{1}\right)\left\{ \left\langle c_{i}\left(t_{2}\right)c_{i}^{\dagger}\left(t_{1}\right)\right\rangle \right.\atop \qquad\qquad\left.-\left\langle c_{i}^{\dagger}\left(t_{1}\right)c_{i}\left(t_{2}\right)\right\rangle \right\} } & \left\langle c_{i}\left(t_{2}\right)c_{i}^{\dagger}\left(t_{1}\right)\right\rangle +\left\langle c_{i}^{\dagger}\left(t_{1}\right)c_{i}\left(t_{2}\right)\right\rangle \end{array}\right)\right\} \\
\equiv\prod_{i}\det\left\{ \mathbb{I}+2\left(\begin{array}{cc}
0 & G_{i}^{R}\\
G_{i}^{A} & G_{i}^{K}\end{array}\right)\left(\begin{array}{cc}
0 & \widetilde{G_{i}^{R}}\\
\widetilde{G_{i}^{A}} & \widetilde{G_{i}^{K}}\end{array}\right)\right\} \end{array}\label{eq:Determinants}\end{equation}

We would like to note the unusual bosonic minus signs in $\widetilde{G_{i}^{R}}\;\&\;\widetilde{G_{i}^{A}}$
in Eq. (\ref{eq:Determinants}) above. The rest of this section is
an evaluation of the determinant in Eq. (\ref{eq:Determinants}) above.
Using the identity $\det\left(\mathbb{I}+M\right)=\exp\left(\sum\frac{-1^{n}}{n}\mathrm{Tr}\left(M^{n}\right)\right)$
we may write that \begin{equation}
\begin{array}{rcl}
\left\langle \gamma\left(0\right)\gamma\left(\mathrm{T}\right)\right\rangle  & = & \exp\left(\sum\frac{-2^{n}}{n}\mathrm{Tr}\left(\sum_{i_{1},i_{2},..i_{2n}}\prod G_{i}^{i_{2k-1},i_{2k}}\widetilde{G_{i}^{i_{2k},i_{2k+1}}}\right)\right)\end{array}\label{eq:Traces}\end{equation}

Here $i_{j}=1\:\mathrm{or}\:2$ and $\left(i_{k},\, i_{k+1}\right)\neq\left(1,\,1\right)$.
To proceed further we will now evaluate each of the traces (to leading
order for large $\mathrm{T}$). As such we need to evaluate integrals
of the form:\begin{eqnarray}
 & \mathrm{\int_{0}^{T}}d\tau_{1}\mathrm{\int_{0}^{T}}d\tau_{2}..\mathrm{\int_{0}^{T}}d\tau_{2n} & \left\{ \left(\left[G_{i}^{A/R/K}\left(\tau_{1}-\tau_{2}\right)\times\left(\theta\left(\tau_{2}-\tau_{1}\right)/\theta\left(\tau_{1}-\tau_{2}\right)/1\right)\right]\times\right.\right.\nonumber \\
 &  & \times\left[G_{i}^{A/R/K}\left(\tau_{3}-\tau_{4}\right)\times\left(\theta\left(\tau_{2}-\tau_{1}\right)/\theta\left(\tau_{1}-\tau_{2}\right)/1\right)\right]\times\label{eq:MultiIntegrals}\\
 &  & \left.....\times\left[G_{i}^{A/R/K}\left(\tau_{2n-1}-\tau_{2n}\right)\times\left(\theta\left(\tau_{2n-1}-\tau_{2n}\right)/\theta\left(\tau_{2n}-\tau_{2n-1}\right)/1\right)\right]\right)\times\nonumber \\
 &  & \times\left(\left[e^{-i\epsilon_{i}\left(\tau_{2}-\tau_{3}\right)-\kappa_{i}\left|\tau_{2}-\tau_{3}\right|}\times\left(\left(1-2n_{i}\right)\theta\left(\tau_{2n-1}-\tau_{2n}\right)/\left(2n_{i}-1\right)\theta\left(\tau_{2n}-\tau_{2n-1}\right)/1\right)\right]\times\right.\nonumber \\
 &  & \left.\left.....\times\left[e^{-i\epsilon_{i}\left(\tau_{2n}-\tau_{1}\right)-\kappa_{i}\left|\tau_{2n}-\tau_{1}\right|}\times\left(\left(1-2n_{i}\right)\theta\left(\tau_{1}-\tau_{2n}\right)/\left(2n_{i}-1\right)\theta\left(\tau_{2n}-\tau_{1}\right)/1\right)\right]\right)\right\} \nonumber \end{eqnarray}

Here for future convenience we have written out the various theta
functions involved and for simplicity assumed relaxation time approximation
for the fermion Greens functions. The terms $A/R/K$ refer to advanced/retarded/Keldysh
Green's functions while the various options for the theta functions
shown in the brackets correspond to the respective green's functions
($A/R/K$). We now need to evaluate these integrals. As a first step
we take advantage of the short range of our correlation functions
(see Fig. (\ref{fig:Correlator})) to change range of integration
limits for the variables $\tau_{1},\tau_{3},...\tau_{2n-1}$ from
$\left(0,\mathrm{T}\right)$ to $\left(-\infty,\infty\right)$. We
also shift the variables of integration calling $u_{i}\equiv\tau_{2i-1}-\tau_{2i},\, v_{i}\equiv\tau_{2i}$.
Combing all these changes we get that the any term in expansion in
Eq. (\ref{eq:Traces}) e.g. Eq. (\ref{eq:MultiIntegrals}) may be
written as:\begin{eqnarray}
\int_{-\infty}^{\infty}du_{1}\int_{-\infty}^{\infty}du_{2}....\int_{-\infty}^{\infty}du_{n} & \times & \left(G_{i}^{A/R/K}\left(u_{1}\right)\times e^{-i\epsilon_{i}u_{1}-\kappa_{i}\left|u_{1}\right|}\times\left(\theta\left(-u_{1}\right)/\theta\left(u_{1}\right)/1\right)\right)\times\label{eq:SimplifiedMultiIntegral}\\
 &  & .......\times\left(G_{i}^{A/R/K}\left(u_{n}\right)\times e^{-i\epsilon_{i}u_{n}-\kappa_{i}\left|u_{n}\right|}\times\left(\theta\left(-u_{n}\right)/\theta\left(u_{n}\right)/1\right)\right)\times\nonumber \\
\times\int_{0}^{T}dv_{1}\int_{0}^{T}dv_{2}....\int_{0}^{T}dv_{n} & \times & \left(\theta\left(v_{2}-v_{1}+u_{2}\right)/\theta\left(v_{1}-v_{2}-u_{2}\right)/1\right).....\left(\theta\left(v_{1}-v_{n}+u_{1}\right)/\theta\left(v_{n}-v_{1}-u_{1}\right)/1\right)\nonumber \end{eqnarray}

We may further simplify this expression by noting that all the correlation
functions $G_{i}^{A/R/K}$ are dominated by small values of $\mathrm{u}$
so that we may approximate $\theta\left(v_{2}-v_{1}+u_{2}\right)\cong\theta\left(v_{2}-v_{1}\right)$
and similarly for other $\theta$ functions. Substituting we get that
the integrals simplify:\begin{equation}
\left\{ \prod_{j=1}^{n}\int_{-\infty}^{\infty}G_{i}^{A/R/K}\left(u_{j}\right)e^{-i\epsilon_{i}u_{j}-\kappa_{i}\left|u_{j}\right|}\cdot\left(\theta\left(-u_{i}\right)/\theta\left(u_{i}\right)/1\right)\right\} \times\left\{ \int_{0}^{T}dv_{1}..\int_{0}^{T}dv_{n}\prod_{j=1}^{n}\left(\theta\left(v_{j+1}-v_{j}\right)/\theta\left(v_{j}-v_{j+1}\right)/1\right)\right\} \label{eq:SimplifedIntegral}\end{equation}

In Appendix \ref{sec:Tedious-Calculations} we will further simplify
the expression in Eq. (\ref{eq:SimplifedIntegral}) above. Here we
will merely compute the leading order term for the semi classical
case where $G_{i}^{K}\gg G_{i}^{R},G_{i}^{A}$. In this case a single
term (containing only $G_{i}^{K}$ contributions) dominates at each
order of integration and we may write that:\begin{equation}
\mathrm{\mathrm{Tr}}\left(\sum_{i_{1},i_{2},..i_{2n}}\prod G_{i}^{i_{2k-1},i_{2k}}\widetilde{G_{i}^{i_{2k},i_{2k+1}}}\right)\cong\left(\widehat{G_{i}^{K}}\left(\epsilon_{i}-i\kappa_{i}\right)\cdot\mathrm{T}\right)^{n}\label{eq:SimplifiedQuantumCoherence}\end{equation}

Here $\widehat{G_{i}^{K}}\left(\epsilon_{i}-i\kappa_{i}\right)$ is
the {}``Fourier transform'' of the Keldysh Green's function evaluated
at energy $\epsilon_{i}$ and decay term $\kappa_{i}$. Combining
these results we recover the semiclassical result that:\begin{eqnarray}
\left\langle \mathrm{\gamma\left(0\right)\gamma\left(T\right)}\right\rangle  & = & \prod_{i}\frac{1}{1+2\mathrm{T}\widehat{G_{i}^{K}}\left(\epsilon_{i}-i\kappa_{i}\right)}\label{eq:KeldishFinalSemiclassical}\\
 & = & \prod_{i}\frac{1}{1+2\mathrm{T}\left(\widehat{G_{i}}\left(\epsilon_{i}-i\kappa_{i}\right)+\widehat{G_{i}}\left(-\epsilon_{i}+i\kappa_{i}\right)\right)}\nonumber \\
 & \cong & \exp\left(-2\mathrm{T}\sum_{i}\left(\widehat{G_{i}}\left(\epsilon_{i}-i\kappa_{i}\right)+\widehat{G_{i}}\left(-\epsilon_{i}+i\kappa_{i}\right)\right)\right)\nonumber \end{eqnarray}

\end{widetext}In the second step we have used a relation between
Keldysh and time ordered correlation functions and in the last step
we have assumed that there are many relevant fermionic modes in the
bath so that we can safely exponentiate each term. Further corrections
to this result are given in Appendix \ref{sec:Tedious-Calculations}.

\section{Various Tedious Calculations and Proofs\label{sec:Tedious-Calculations}}

\subsection{\label{sub:Parity-Eigenvalues-(Coding}Parity eigenvalues (coding
subspace) }

In the main text (see Section \ref{sec:Introduction}) we presented
a specific encoding of the Majorana qubit that used the even Majorana
fermion parity subspace for its coding space. Throughout the main
text we computed expectation values of the form $\mathrm{\left\langle \gamma_{1}\left(0\right)\gamma_{2}\left(0\right)\gamma_{1}\left(T\right)\gamma_{2}\left(T\right)\right\rangle =-\left\langle \sigma^{z}\left(0\right)\sigma^{z}\left(T\right)\right\rangle }$.
We claimed that this is a good representation of the fidelity of our
quantum memory. There could be further concern that we are over or
under estimating the fidelity by including in the expectation value
$\left\langle \mathrm{\gamma(0)\gamma...\gamma(T)}\right\rangle $
processes that included final states that do not have an even fermion
parity \cite{key-45}. Here we show that for two time correlation
functions such processes never contribute to this expectation value
so no further measurements or corrections are needed to adjust for
such processes. Even though we do not focus on this case in the main
text we will show that the above statement is not correct for multitime
correlators. We will also show what modifications must be made in
the multitime case.

\subsubsection{\label{sub:Two-Time-Correlators}Two time correlators}

We start by showing that no modifications are necessary in the two
time correlators case (again focusing on the four Majorana fermion
qubit). Indeed consider $\prod_{+}$ and $\prod_{-}$ projectors into
even and odd Majorana fermion parity subspaces ($\prod_{+}+\prod_{-}=1$,
$\prod_{\pm}^{2}=\prod_{\pm}$ and $\prod_{+}\prod_{-}=0$). Since
the initial state of the Majorana qubit has even fermion parity, we
may write that:\begin{equation}
\begin{array}{l}
\left\langle \mathrm{\sigma^{\mathrm{z}}\left(0\right)\sigma^{\mathrm{z}}\left(T\right)}\right\rangle =\left\langle \mathrm{\prod_{+}\sigma^{\mathrm{z}}\left(0\right)\sigma^{\mathrm{z}}\left(T\right)\prod_{+}}\right\rangle \\
=\left\langle \mathrm{\prod_{+}\sigma^{\mathrm{z}}\left(0\right)\left(\prod_{+}+\prod_{-}\right)\sigma^{\mathrm{z}}\left(T\right)\prod_{+}}\right\rangle \\
=\left\langle \prod_{+}\mathrm{\sigma^{\mathrm{z}}\left(0\right)\prod_{+}\sigma^{\mathrm{z}}\left(T\right)\prod_{+}}\right\rangle \\
=\left\langle \mathrm{\sigma^{\mathrm{z}}\left(0\right)\prod_{+}\sigma^{\mathrm{z}}\left(T\right)\prod_{+}}\right\rangle \end{array}\label{eq:Correlators}\end{equation}
 In the third step we have used the fact that $\left[\mathrm{\sigma^{z}}\left(0\right),\:\prod_{\pm}\right]=0$
to get rid of the term $\prod_{+}\sigma^{z}\left(0\right)\prod_{-}=0$.
From this we see that we may as well project out the odd fermion parity
subspace, e.g. $\mathrm{\mathrm{\sigma^{z}}\left(T\right)\rightarrow\prod_{+}\mathrm{\sigma^{z}}\left(T\right)}\prod_{+}$
and not worry about errors involving non-coding subspaces (these errors
do not contribute to expectation values). The same sort of argument
may be made for any two time correlator of the fermion modes and any
encoding subspace. Indeed based on the form of the previous proof
to ensure that the non-coding subspace does not contribute to the
expectation values all we need is a coding system such that the logic
operators do not take us out of the encoding space (which is always
the case). So no further corrections are needed in this case.

\subsubsection{\label{sub:Multi-Time-Correlators}Multi-time correlators}

In the multi time case in order to only consider terms within the
even fermion parity subspace it is necessary to project out the odd
fermion parity states explicitly; that is convert $\mathcal{O}_{i}\left(\mathrm{T}\right)\rightarrow\prod_{+}\mathcal{O}_{i}\left(\mathrm{T}\right)\prod_{+}$.
There are still many simplifications in the case of three time correlations.
In this case similarly to what we did above one can check that it
is only necessary to project out once just before the last operator.
For example:\begin{equation}
\begin{array}{l}
\left\langle \sigma^{\mathrm{z}}\left(0\right)\sigma^{\mathrm{z}}\left(\tau_{1}\right)\sigma^{\mathrm{z}}\left(\tau_{2}\right)\right\rangle \rightarrow\left\langle \sigma^{\mathrm{z}}\left(0\right)\sigma^{\mathrm{z}}\left(\tau_{1}\right)\prod_{+}\sigma^{\mathrm{z}}\left(\tau_{2}\right)\right\rangle =\\
\mathrm{\frac{-i}{2}}\left\langle \gamma_{1}\gamma_{2}\gamma_{1}\left(\tau_{1}\right)\gamma_{2}\left(\tau_{1}\right)\left(1+\gamma_{1}\gamma_{2}\gamma_{3}\gamma_{4}\right)\gamma_{1}\left(\tau_{2}\right)\gamma_{2}\left(\tau_{2}\right)\right\rangle ,\end{array}\label{eq:Multitime}\end{equation}
 which we can calculate using the methods derived in this paper.

\subsection{\label{sub:Cross-Correlations-between}Cross Correlations between
Majorana baths}

In the bulk of the text we have discussed the case when the different
baths surrounding the Majorana fermions are uncorrelated, or equivalently
that interactions between modes that couple to different Majorana
fermions are negligible. In this section we shall discuss the effects
of such interactions, and indeed argue that they may well be neglected
in the case of well separated Majorana modes: modes whose separation
is much greater then the scattering length in the bath medium.

First we begin by arguing that the initial conditions which we have
selected in this paper, of uncorrelated distant baths, are likely
to be highly favorable for the coherence of a qubit composed of Majorana
fermions. Indeed, focusing on two Majorana modes, we note that the
coherence of the qubit may be expressed as $\left\langle \gamma_{1}\gamma_{2}\; e^{i{H}\mathrm{T}}\;\gamma_{1}\gamma_{2}\; e^{-i{H}\mathrm{T}}\right\rangle $.
We now consider two Majorana modes each interacting with the same
fermionic environment: in particular we will focus on a shared modes
$f_{\epsilon}$ with energy $\epsilon$, coupling to both $\gamma_{1}$
and $\gamma_{2}$ through a Hamiltonian of the form $H=\gamma_{1}\sum_{\epsilon}\left(\Gamma_{1}^{\epsilon}\, f_{\epsilon}-{\Gamma_{1}^{\epsilon}}^{*}\, f_{\epsilon}^{\dagger}\right)+\gamma_{2}\sum_{\epsilon}\left(\Gamma_{2}^{\epsilon}\, f_{\epsilon}-{\Gamma_{2}^{\epsilon}}^{*}\, f_{\epsilon}^{\dagger}\right)$.
Here $\Gamma_{1,2}^{\epsilon}$ are just complex tunneling amplitudes,
for simplicity. Taylor expanding the exponentials in the equation
above, we obtain non-zero contributions to the coherence (the expectation
value given above) that contain cross terms involving both of $\Gamma_{1}^{\epsilon}$
and $\Gamma_{2}^{\epsilon}$: \begin{widetext} \begin{eqnarray}
 &  & -2\left\langle \gamma_{1}\gamma_{2}\right\rangle \,\int_{0}^{\mathrm{T}}\!\!\! dt_{1}\!\int_{0}^{\mathrm{T}}\!\!\! dt_{2}\;\sum_{\epsilon}\;\langle\left[\left(\Gamma_{1}^{\epsilon}\, f_{\epsilon}(t_{1})-{\Gamma_{1}^{\epsilon}}^{*}\, f_{\epsilon}^{\dagger}(t_{1})\right),\left(\Gamma_{2}^{\epsilon}\, f_{\epsilon}(t_{2})-{\Gamma_{2}^{\epsilon}}^{*}\, f_{\epsilon}^{\dagger}(t_{2})\right)\right]\rangle\nonumber \\
 &  & =2\left\langle \gamma_{1}\gamma_{2}\right\rangle \,\int_{0}^{\mathrm{T}}\!\!\! dt_{1}\!\int_{0}^{\mathrm{T}}\!\!\! dt_{2}\;\sum_{\epsilon}\;{\Gamma_{1}^{\epsilon}}^{*}\Gamma_{2}^{\epsilon}\;\left(\langle f_{\epsilon}(t_{2})f_{\epsilon}^{\dagger}(t_{1})\rangle-\langle f_{\epsilon}^{\dagger}(t_{1})f_{\epsilon}(t_{2})\rangle\right)+{\rm h.c.}\label{eq:SingleBathEquation}\end{eqnarray}
 \end{widetext} These are the interference terms that do not appear
for Majorana fermions interacting with separate baths, but appear
due to a common bath. For short times any non-zero terms like those
lead to decoherence. Indeed, since it is impossible to have higher
then unity coherence, these terms must contribute negatively to the
performance of a qubit composed of Majorana fermions.

However we would like to now argue that this effect can easily be
avoided in realistic experimental situations by simply keeping the
Majorana fermions far apart. First note that individual $f$ modes
that are localized cannot have large tunneling overlaps with two distant
Majoranas, so $\Gamma_{1}\Gamma_{2}^{*}\cong0$. Therefore only extended
modes can contribute to the interference terms. Now, each such mode
contains a normalization factor proportional to inverse square root
of volume, so individually they contribute zero in the thermodynamic
limit. As such, in order to get a non-zero value for the term shown
in Eq. (\ref{eq:SingleBathEquation}) we need to integrate over the
contributions of all the extended states. To do so first recall Eq.
(\ref{eq:TunnelingHamiltonian}) or Eq. (\ref{eq:Tunneling}) below
which state that $\Gamma_{1,2}^{\epsilon}\sim\int dr\; u_{1,2}\left(r\right)\times v_{\epsilon}\left(r\right)$.
Here $u_{1,2}$ is the wavefunction of the Majorana mode while $v_{\epsilon}$
is the wavefunction of the mode $f_{\epsilon}$. Assuming a pointlike
$u_{1,2}$ or dividing the integral into portions of negligible extent
we may write that $\Gamma_{1,2}^{\epsilon}\propto v_{\epsilon}\left(r_{1,2}\right)$,
where $r_{1,2}$ are the locations of the two Majorana modes. In this
case, we can relate terms entering Eq. (\ref{eq:SingleBathEquation})
to single-particle Green's functions for the bath electrons: \begin{eqnarray}
 &  & \sum_{\epsilon}{\Gamma_{1}^{\epsilon}}^{*}\Gamma_{2}^{\epsilon}\;\langle f_{\epsilon}(t_{2})f_{\epsilon}^{\dagger}(t_{1})\rangle\nonumber \\
 &  & \propto\sum_{\epsilon}v_{\epsilon}^{*}\left(r_{1}\right)v_{\epsilon}\left(r_{2}\right)\;\langle f_{\epsilon}(t_{2})f_{\epsilon}^{\dagger}(t_{1})\rangle\nonumber \\
 &  & =G\left(r_{1}\,,\, t_{1}\,;\, r_{2}\,,\, t_{2}\right)\;.\end{eqnarray}
 In a realistic material there are always sources of decorrelation,
in particular lattice disorder and phonons. It is not too difficult
to show that\cite{key-12,key-13,key-19} these sources lead to an
exponential decay of $G\left(r_{1}\,,\, t_{1}\,;\, r_{2}\,,\, t_{2}\right)$
in space with a characteristic length given by the mean free path
of the material. The mean free path is directly related to phonon
and impurity scattering strengths\cite{key-12,key-13,key-19}. Since
this reasoning indicates an exponential suppression of these interference
effects with distance, and since it is not possible to use these interference
effects to enhance coherence anyway, we have ignored the possibility
of the Majorana modes sharing a common bath in the text.

\subsection{Partial justification of independently fluctuating modes.\label{sec:Justification}}

In Section \ref{sec:Fluctuating-Hamiltonians} we presented some results
for the coherence of a single Majorana mode in the presence of a fluctuating
environment. While we covered both diagonal fluctuations and cross
correlations between different modes of our environment, we mostly
focused on the case of diagonal fluctuations. Furthermore our results
on cross-correlations are technical and in practice difficult to apply.
Here we shall present a partial justification indicating that diagonal
fluctuations are dominant over cross correlations. Weak correlations
do exist so no {}``theorem'' indicating a lack of cross-correlations
can be presented. We will however present arguments supporting independent
correlations in three key cases: when there is a high degree of symmetry
for the problem, when there is {}``disorder averaging'' of the continuum
states and tunnel couplings have short correlation length, or to leading
order in perturbation when the fluctuations are weak.

\subsubsection{\label{sub:High-degree-of}High degree of symmetry}

Many Hamiltonians have a high degree of symmetry. For example for
a p-wave superconductor with a single vortex supporting a single Majorana
mode the vortex core states have rotational symmetry. Most external
Hamiltonians causing fluctuations in the vortex core are invariant
under this rotational symmetry and as such they may be written in
block diagonal form with each block corresponding to a different eigenstate
of the rotation operator. As such fluctuations corresponding to different
angular momentum eigenstates are decoupled from each other (uncorrelated),
justifying this assumption in this case. More generally fermionic
modes corresponding to different irreducible representations (diagonal
blocks) of some fluctuation Hamiltonian have uncorrelated fluctuations.
This in part justifies the assumptions used in Section \ref{sec:Fluctuating-Hamiltonians}.

\subsubsection{\label{sub:Long-Wavelength-Fluctuations}Short correlation length
\& disorder averaging}

We shall now focus on a particularly simple, but realistic, model
of tunnel couplings between the Majorana mode and the regular fermion
modes in the superconductor. We shall assume point like tunneling
with an effective coupling that may be written as: \begin{widetext}\begin{eqnarray}
H_{\mathrm{tun}} & = & \gamma\sum_{i}\left\{ c_{i}\left(\int d^{2}r\left\{ \Xi\left(r,\,\tau\right)u_{0}\left(r\right)u_{i}\left(r\right)-\Xi^{*}\left(r,\,\tau\right)v_{0}\left(r\right)v_{i}\left(r\right)\right\} \right)\right.\nonumber \\
 & + & \left.c_{i}^{\dagger}\left(\int d^{2}r\left\{ \Xi\left(r,\,\tau\right)u_{0}\left(r\right)v_{i}^{*}\left(r\right)-\Xi^{*}\left(r,\,\tau\right)v_{0}\left(r\right)u_{i}^{*}\left(r\right)\right\} \right)\right\} .\label{eq:TunnelingHamiltonianModel}\end{eqnarray}

Here $u_{i}\left(r\right)$ and $v_{i}\left(r\right)$ are the creation
and annihilation components of the modes $c_{i}$ while $u_{0}\left(r\right)$
and $v_{0}\left(r\right)$ are the creation and annihilation components
of the Majorana mode and $\Xi$ is a tunneling amplitude. For a similar
coupling form see e.g. Eqs. (\ref{eq:BasicTunnelingHamiltonian}),
\& (\ref{eq:TunnelingHamiltonian}). From this we see that within
our model the coupling functions in Eq. (\ref{eq:Main Hamiltonian})
is given by:\begin{equation}
\Gamma_{i}\left(\tau\right)=\int d^{2}r\left\{ \Xi\left(r,\,\tau\right)u_{0}\left(r\right)u_{i}\left(r\right)-\Xi^{*}\left(r,\,\tau\right)v_{0}\left(r\right)v_{i}\left(r\right)\right\} .\label{eq:Tunneling}\end{equation}

The correlation function is given by:\begin{eqnarray}
\left\langle \Gamma_{i}^{*}\left(\tau_{1}\right)\Gamma_{j}\left(\tau_{2}\right)\right\rangle  & = & -\int d^{2}r_{1}\int d^{2}r_{2}\left\langle \Xi\left(r_{1},\,\tau_{1}\right)\Xi^{*}\left(r_{2},\,\tau_{2}\right)u_{0}\left(r_{1}\right)v_{0}\left(r_{2}\right)u_{i}\left(r_{1}\right)u_{j}^{*}\left(r_{2}\right)\:+\right.\nonumber \\
 &  & \qquad\qquad\qquad\qquad\left.+\;\Xi\left(r_{1},\,\tau_{1}\right)\Xi^{*}\left(r_{2},\,\tau_{2}\right)u_{0}\left(r_{1}\right)v_{0}\left(r_{2}\right)v_{i}\left(r_{1}\right)v_{j}^{*}\left(r_{2}\right)\right\rangle \nonumber \\
 & \cong & -\int d^{2}r\left\{ F\left(\tau_{1},\:\tau_{2}\right)\left\langle \left|u_{0}\left(r\right)\right|^{2}u_{i}\left(r\right)u_{j}^{*}\left(r\right)\right\rangle \,+\, F^{*}\left(\tau_{1},\:\tau_{2}\right)\left\langle \left|u_{0}\left(r\right)\right|^{2}v_{i}\left(r\right)v_{j}^{*}\left(r\right)\right\rangle \right\} \nonumber \\
 & \cong & -\int d^{2}r\left\{ F\left(\tau_{1},\:\tau_{2}\right)\left\langle \left|u_{0}\left(r\right)\right|^{2}\mathcal{U}_{i}\left(r\right)\delta_{ij}\right\rangle \,+\, F^{*}\left(\tau_{1},\:\tau_{2}\right)\left\langle \left|u_{0}\left(r\right)\right|^{2}\mathcal{V}_{i}\left(r\right)\delta_{ij}\right\rangle \right\} .\label{eq:CrossFluctuations}\end{eqnarray}

\end{widetext}Here we able to simplify our expressions by assuming
that $\left\langle \Xi\left(\vec{r}_{1},\,\tau_{1}\right)\Xi^{*}\left(\vec{r}_{2},\,\tau_{2}\right)\right\rangle \cong F\left(\tau_{1},\:\tau_{2}\right)\delta\left(\vec{r}_{1}-\vec{r_{2}}\right)$
for some $F\left(\tau_{1},\:\tau_{2}\right)$ and that $\left\langle \Xi\left(\vec{r}_{1},\,\tau_{1}\right)\Xi^{*}\left(\vec{r}_{2},\,\tau_{2}\right)\right\rangle \cong0$.
We have also performed a disorder average over the bath states $u_{i}\left(r\right)u_{j}\left(r\right)\sim\delta_{ij}$.
This averaging works well for continuum states.

\subsubsection{\label{sub:Weak Disorder}Weak Fluctuations}

In many situations there are many fermionic modes responsible for
the decoherence of the Majorana mode and the coupling to any one mode
is quite weak. In this case even if the fluctuations between the different
fermion modes are strongly cross correlated the diagonal correlations
dominate decoherence. Indeed, to show this we first recall the formula
for the coherence of a Majorana correlator given in Section \ref{sec:Cross-Correlations}:
$\left\langle \gamma\left(0\right)\gamma\left(\mathrm{T}\right)\right\rangle =\mathrm{det}^{-1}\left(\mathbb{I}+2\boldsymbol{\mathbf{\sigma}}\left(\mathrm{T}\right)\right)$.
We now simplify this formula. First, letting the eigenvalues of $\boldsymbol{\sigma}$
be $\left\{ \lambda_{i}\right\} $, we obtain that:\begin{eqnarray}
\left\langle \gamma\left(0\right)\gamma\left(\mathrm{T}\right)\right\rangle  & = & \prod_{i}\frac{1}{1+2\lambda_{i}}\label{eq:EigenvaluesTrace}\\
 & \cong & \exp\left(-2\sum\lambda_{i}\right)=\exp\left(-2\mathrm{Tr}\left(\mathbf{\boldsymbol{\sigma}}\right)\right)\nonumber \end{eqnarray}
 In the second step we have assumed that many eigenvalues contribute
to the product so we can exponentiate. From this we see explicitly
that in many cases with weak fluctuations only diagonal terms of the
matrix $\sigma$ matter. These are one particle terms $\boldsymbol{\sigma_{ii}}\left(\mathrm{T}\right)\equiv2\int_{0}^{\mathrm{T}}\; d\tau_{1}\;\int_{0}^{\mathrm{T}}\; d\tau_{2}\;\; e^{-i\epsilon_{i}\tau_{1}}\; G_{i}(\tau_{1},\tau_{2})\; e^{+i\epsilon_{i}\tau_{2}}$
and as such are much easier to handle.

\subsection{\label{sub:Proofs-and-Clarifications}Proofs and clarifications of
Eqs. (\ref{eq:Determinant}), (\ref{eq:SimplifedIntegral}), \& (\ref{eq:telegraph-noise})}

\subsubsection{\label{sub:Eq.-Determinant.}Eq. (\ref{eq:Determinant}).}

Here we wish to prove Eq. (\ref{eq:Determinant}) for arbitrary (not
necessarily Hermitian) matrices. As a first step we wish to prove
an analogous expression for real Gaussian integrals. More precisely
we wish to show that for an arbitrary possibly complex $n\times n$
matrix $M$ and an integral over $\mathbb{R}^{n}$ we may write that:\begin{equation}
\int dx_{1}...dx_{n}\exp\left(-\frac{1}{2}\vec{x}^{T}M\vec{x}\right)=\frac{\left(2\pi\right)^{n/2}}{\left(\det\left(\frac{M+M^{T}}{2}\right)\right)^{1/2}}\label{eq:RealDeterminant}\end{equation}

To prove this we first note that $\sum_{i,j}x_{i}M_{ij}x_{j}=\frac{1}{2}\sum x_{i}\left(M_{ij}+M_{ji}\right)x_{j}$.
As such we may safely transform $M\rightarrow\frac{1}{2}\left(M+M^{T}\right)$.
Next we may use Takagi's decomposition for symmetric matrices \cite{key-47}
to write that $\frac{1}{2}\left(M+M^{T}\right)=UDU^{T}$. Where $U$
is a unitary matrix and $D$ is a diagonal one. From this we see that\begin{widetext}
\begin{equation}
\int dx_{1}...dx_{n}\exp\left(-\frac{1}{2}\vec{x}^{T}M\vec{x}\right)=\frac{\left(2\pi\right)^{n/2}}{\left(\det\left(D\right)\right)^{1/2}\det\left(U\right)}=\frac{\left(2\pi\right)^{n/2}}{\left(\det\left(\frac{1}{2}\left(M+M^{T}\right)\right)\right)^{1/2}}\label{eq:RealTrace}\end{equation}
 The extra factor of $\det\left(U\right)$ comes from the Jacobian
of the change of variables. To proceed to the complex case we begin
by writing $\vec{z}=\vec{x}+i\vec{y},\,\vec{z}^{*}=\vec{x}-i\vec{y}$.
Then we may write that:\begin{equation}
\vec{z}^{\dagger}G^{-1}\vec{z}=\left(\begin{array}{cc}
\vec{x}^{T} & \vec{y}^{T}\end{array}\right)\left(\begin{array}{cc}
G^{-1} & iG^{-1}\\
-iG^{-1} & G^{-1}\end{array}\right)\left(\begin{array}{c}
\vec{x}\\
\vec{y}\end{array}\right)\label{eq:RealPart}\end{equation}

As such we may write that:\begin{eqnarray}
\int\int dz_{1}..dz_{n}dz_{1}^{*}..dz_{n}^{*}\exp\left(-\frac{1}{2}\vec{z}^{\dagger}G^{-1}\vec{z}\right) & = & \int\int dx_{1}..dx_{n}dy_{1}..dy_{n}\exp\left(-\frac{1}{2}\left(\begin{array}{cc}
\vec{x}^{T} & \vec{y}^{T}\end{array}\right)\left(\begin{array}{cc}
G^{-1} & iG^{-1}\\
-iG^{-1} & G^{-1}\end{array}\right)\left(\begin{array}{c}
\vec{x}\\
\vec{y}\end{array}\right)\right)\nonumber \\
 & = & \left(2\pi\right)^{n}\left(\det\left[\frac{1}{2}\left(\left(\begin{array}{cc}
G^{-1} & iG^{-1}\\
-iG^{-1} & G^{-1}\end{array}\right)+\left(\begin{array}{cc}
G^{-1} & iG^{-1}\\
-iG^{-1} & G^{-1}\end{array}\right)^{T}\right)\right]\right)^{-\frac{1}{2}}\label{eq:SymmetricComplex}\end{eqnarray}

Next we note that:\begin{equation}
\frac{1}{2}\left(\begin{array}{cc}
G^{-1}+G^{-1T} & i\left(G^{-1}-G^{-1T}\right)\\
-i\left(G^{-1}-G^{-1T}\right) & G^{-1}+G^{-1T}\end{array}\right)=\left(\begin{array}{cc}
1 & i\\
0 & 1\end{array}\right)\left(\begin{array}{cc}
G^{-1} & 0\\
\frac{-i}{2}\left(G^{-1}-G^{-1T}\right) & G^{-1T}\end{array}\right)\left(\begin{array}{cc}
1 & -i\\
0 & 1\end{array}\right)\label{eq:Transformation}\end{equation}

Since \begin{equation}
\det\left(\begin{array}{cc}
1 & -i\\
0 & 1\end{array}\right)=\det\left(\begin{array}{cc}
1 & i\\
0 & 1\end{array}\right)=1,\,\det\left(\begin{array}{cc}
G^{-1} & 0\\
\frac{-i}{2}\left(G^{-1}-G^{-1T}\right) & G^{-1T}\end{array}\right)=\det\left(G^{-1}\right)\det\left(G^{-1T}\right)\label{eq:DeterminantProducts}\end{equation}

We get that\begin{equation}
\int\int dz_{1}...dz_{n}dz_{1}^{*}...dz_{n}^{*}\exp\left(-\frac{1}{2}\vec{z}^{\dagger}G^{-1}\vec{z}\right)=\left(2\pi\right)^{n}\left(\det\left(G^{-1}\right)\det\left(G^{-1T}\right)\right)^{-1/2}=\left(2\pi\right)^{n}\det\left(G\right)\label{eq:FinalAnswerDeterminant}\end{equation}
 \end{widetext}This reproduces Eq. (\ref{eq:Determinant}).

\subsubsection{\label{sub:Eq.-SimplifiedIntegral}Eq. (\ref{eq:SimplifedIntegral})}

\begin{figure}
\begin{centering}
\includegraphics[scale=0.9]{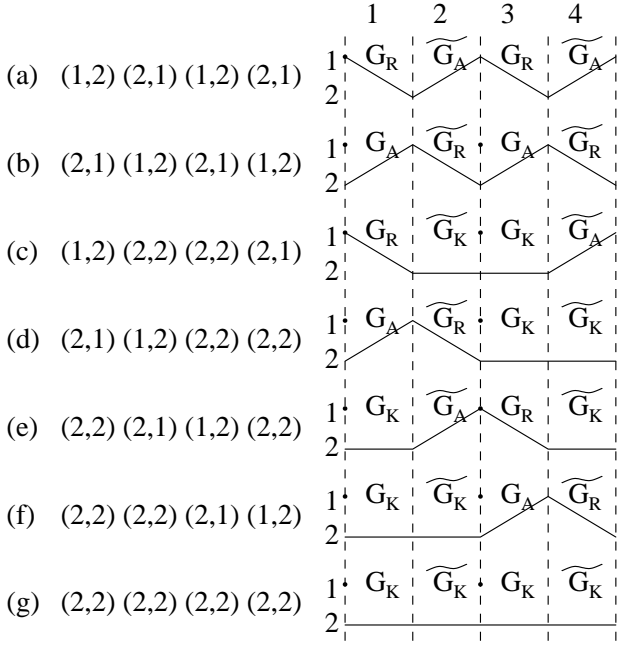} 
\par\end{centering}

\caption{\label{cap:TraceKeldish} In this figure we consider the second order
term in Eq. (\ref{eq:SimplifedIntegral}) above. We picture the seven
terms contributing to $\mathrm{Tr\left\{ \left(\left(\protect\begin{array}{cc}
0 & G_{i}^{R}\protect\\
G_{i}^{A} & G_{i}^{K}\protect\end{array}\right)\left(\protect\begin{array}{cc}
0 & \widetilde{G_{i}^{R}}\protect\\
\widetilde{G_{i}^{A}} & \widetilde{G_{i}^{K}}\protect\end{array}\right)\right)^{2}\right\} }$with lines connecting indices in the Keldysh matrix, e.g. $\left(1,\:2\right)$
stands for $G^{\left(1,2\right)}=G^{R}$. Each entry corresponds to
a Green's function. The biggest term contains four Keldysh Green's
functions (pictured last (g)). The six subleading terms are also shown.}

\end{figure}

Here we would like to further simplify the sums in Eqs. (\ref{eq:SimplifedIntegral})
and (\ref{eq:Traces}) as well as obtain more accurate estimates.
We begin with Eq. (\ref{eq:SimplifedIntegral}) above. By considering
the form of the indices in the trace we see that we may represent
any term in the expansion for $\mathrm{Tr}\left\{ \left(\left(\begin{array}{cc}
0 & G_{i}^{R}\\
G_{i}^{A} & G_{i}^{K}\end{array}\right)\left(\begin{array}{cc}
0 & \widetilde{G_{i}^{R}}\\
\widetilde{G_{i}^{A}} & \widetilde{G_{i}^{K}}\end{array}\right)\right)^{n}\right\} $ as a set of broken lines with periodic boundary conditions with each
line representing an appropriate Green's function (see Fig. (\ref{cap:TraceKeldish})).
In the quasi classical limit the biggest contribution comes from the
term $Tr\left\{ \left(G_{i}^{K}\widetilde{G_{i}^{K}}\right)^{n}\right\} \simeq\mathrm{T}^{n}\left(G_{i}^{K}\left(\epsilon_{i}-i\kappa_{i}\right)\right)^{n}$.
The last equality may be obtained by noting that the various terms
in Eq. (\ref{eq:SimplifedIntegral}) factorize. By noting that most
of Eq. (\ref{eq:SimplifedIntegral}) factorizes we may compute the
subleading term including combinatorial factors in the semiclassical
expansion, it is $\frac{n}{4}\mathrm{T}^{n}\left(\widehat{G_{i}^{K}}\left(\epsilon-i\kappa_{i}\right)\right)^{n-1}\left(\widehat{G_{i}^{R}}\left(\epsilon-i\kappa_{i}\right)+\widehat{G_{i}^{A}}\left(\epsilon+i\kappa_{i}\right)\right)$
(for $\mathrm{n\geq1}$). This term would correspond to diagrams (c)-(f)
in Fig. (\ref{cap:TraceKeldish}). As such we obtain that:\begin{widetext}\begin{eqnarray}
\left\langle \gamma\left(0\right)\gamma\left(\mathrm{T}\right)\right\rangle  & \cong & \prod_{i}\exp\left\{ \sum_{n=0}^{\infty}\frac{\left(-2\right)^{n}}{n}\mathrm{T}^{n}\left(\widehat{G_{i}^{K}}\left(\epsilon_{i}-i\kappa_{i}\right)\right)^{n}+\right.\nonumber \\
 & + & \left.\sum_{n=1}^{\infty}\frac{\left(-2\right)^{n}}{4}\mathrm{T}^{n}\left(\widehat{G_{i}^{K}}\left(\epsilon_{i}-i\kappa_{i}\right)\right)^{n-1}\left(\widehat{G_{i}^{R}}\left(\epsilon_{i}-i\kappa_{i}\right)+\widehat{G_{i}^{A}}\left(\epsilon_{i}+i\kappa_{i}\right)\right)\right\} \nonumber \\
 & \cong & \prod_{i}\frac{1}{1+2\mathrm{T}\widehat{G_{i}^{K}}\left(\epsilon_{i}-i\kappa_{i}\right)}\exp\left(-\frac{1}{2}\mathrm{T}\left(\widehat{G_{i}^{R}}\left(\epsilon_{i}-i\kappa_{1}\right)+\widehat{G_{i}^{A}}\left(\epsilon_{i}+i\kappa_{i}\right)\right)\cdot\frac{1}{1+2\mathrm{T}\widehat{G_{i}^{K}}\left(\epsilon_{i}-i\kappa_{i}\right)}\right)\nonumber \\
 & \cong & \left[\prod_{i}\exp\left(-2\mathrm{T}\widehat{G_{i}^{K}}\left(\epsilon_{i}-i\kappa_{i}\right)\right)\right]\times\left[\prod_{i}\exp\left(-\frac{\widehat{G_{i}^{R}}\left(\epsilon_{i}-i\kappa_{i}\right)+\widehat{G_{i}^{A}}\left(\epsilon_{i}+i\kappa_{i}\right)}{4\widehat{G_{i}^{K}}\left(\epsilon_{i}-i\kappa_{i}\right)}\right)\right]\label{eq:FinalAnswer}\end{eqnarray}
 \end{widetext}In the final step we have taken the large $\mathrm{T}$
limit. As such we recover the semiclassical approximation and the
leading order quantum correction.

\subsubsection{\label{sub:Eq.-TelegraphNoise}Eq. (\ref{eq:telegraph-noise})}

We would like to derive Eq. (\ref{eq:telegraph-noise}). As a first
step we will calculate the n-point correlation function for telegraphic
noise. We will find that it is short ranged and this will allow us
to calculate the distribution of the {}``displacement'' field $Z_{i}\left(\mathrm{T}\right)$
(see Eq. (\ref{eq:Displacement})) within the dipole approximation.
We will find that the distribution is Gaussian at which point Eq.
(\ref{eq:telegraph-noise}) will follow. First we motivate the dipole
approximation used in Section \ref{sub:Telegraph-Noise}. To do so
we compute the n-point correlation function for tunneling amplitudes
acted on by telegraph noise and observe that it is exponentially short
ranged. That is we extend Eqs. (\ref{eq:TelegraphNoiseTwoPoint})
\& (\ref{eq:TelegraphNoiseflips}) from the main text by showing that
for the i'th mode, $t_{1}<t_{2}<...<t_{N}$, and for $N$ even \cite{key-51}:\begin{equation}
\left\langle \prod_{j=1}^{N}\Gamma\left(t_{j}\right)\right\rangle =\Lambda_{i}^{N}\exp\left(-\frac{2}{\tau_{i}}\sum_{j=1}^{N}\left(t_{2j}-t_{2j-1}\right)\right).\label{eq:TelegraphNoiseNpoint}\end{equation}
 To do so we first we recall the result that for telegraph noise the
probability of having exactly $K$ flips in some set of interval whose
total length in $L$ is given by $\frac{1}{K!}\left(\frac{L}{\tau_{i}}\right)^{K}\exp\left(-\frac{L}{\tau_{i}}\right)$
\cite{key-49}. Now we know that $\Pi_{i=1}^{N}\Gamma\left(\tau_{i}\right)=\pm\Lambda_{i}^{N}$
depending on whether an odd or an even number of the $\Gamma\left(\tau_{i}\right)=-\Lambda$.
At this point it is a straightforward combinatorial argument to show
that:\begin{equation}
\begin{array}{l}
\left\{ \#\Gamma\left(\tau_{i}\right)=-\Lambda\right\} =\left\{ \sum_{j=1}^{N}\#\:\mathrm{Flips\: in}\:\left[t_{2j-1},t_{2j}\right]\right\} \\
\qquad\qquad\qquad\qquad\qquad\qquad\qquad\qquad(\mathrm{\mathrm{mod\,2}})\end{array}\label{eq:Flips}\end{equation}
 Combing these results we get that:\begin{equation}
\begin{array}{l}
\left\langle \prod_{j=1}^{N}\Gamma\left(t_{j}\right)\right\rangle =\sum_{n=0}^{\infty}\left(-1\right)^{n}\frac{1}{n!}\left(\frac{L}{\tau_{i}}\right)^{n}\exp\left(-\frac{L}{\tau_{i}}\right)\\
\qquad\qquad\qquad\quad=\exp\left(-2\frac{L}{\tau_{i}}\right)\end{array}\label{eq:ExpectationPoisson}\end{equation}

Here $L=\sum_{j=1}^{N}\left(t_{2j}-t_{2j-1}\right)$. As such we obtain
the result in Eq. (\ref{eq:TelegraphNoiseNpoint}). Now we wish to
calculate $2n$ point function of the displacement field, see Eq.
(\ref{eq:Displacement}). It is given by:\begin{widetext}\begin{eqnarray}
\left\langle \left|Z_{i}\left(\mathrm{T}\right)\right|^{2n}\right\rangle  & = & 2^{2n}\mathrm{\int}\mathcal{D}\left\{ \Gamma_{i}\left(\tau_{1}\right)\right\} P\left\{ \Gamma_{i}\left(\tau\right)\right\} \int_{0}^{T}d\tau_{1}...\int_{0}^{T}d\tau_{2n}\prod_{i}\exp\left(\vartheta_{k}i\epsilon_{i}\tau_{k}\right)\left\langle \prod_{k}\Gamma\left(\tau_{k}\right)\right\rangle \nonumber \\
 & = & \left(2\Lambda\right)^{2n}\times\mathrm{lim_{\delta\rightarrow0}}\sum_{P_{2n}}\left\{ \sum_{l=0}^{2n}\left\{ \left(-1\right)^{2n-l}\exp\left(\sum_{j=1}^{l}\left\{ \left(\vartheta_{P_{2n}\left(j\right)}i\epsilon_{i}+\delta\right)+2\left(-1\right)^{j}\Omega_{i}\right\} \mathrm{T}\right)\right\} \times\right.\label{eq:Cumulants}\\
 & \times & \left.\left(\prod_{j=1}^{l}\frac{1}{\sum_{k=j}^{l}\left(\vartheta_{P_{2n}\left(k\right)}i\epsilon_{i}+\delta+2\left(-1\right)^{k}\Omega_{i}\right)}\right)\times\left(\prod_{j=l+1}^{2n}\frac{1}{\sum_{k=l+1}^{j}\left(\vartheta_{P_{2n}\left(k\right)}i\epsilon_{i}+\delta+2\left(-1\right)^{k}\Omega_{i}\right)}\right)\right\} \nonumber \end{eqnarray}

\end{widetext} Here $\mathrm{\left\{ \Gamma_{i}\left(\tau\right)\right\} }$
refers to the space of all path alternating between $+\Lambda_{i}$
and $-\Lambda_{i}$ and $\mathrm{P\left\{ \Gamma_{i}\left(\tau\right)\right\} }$
is the probability of such a path, and we have introduced $\vartheta_{k}=\left\{ \begin{array}{l}
1,\:\:\; k\leq n\\
-1,\; k>n\end{array}\right.$. We will derive the second part of this equation separately below.
The limit: $\mathrm{lim_{\delta\rightarrow0}}$ comes from the fact
that some of the denominators may turn to zero without an extra factor
of $\delta$. Also we would like to note that there is a sum over
the permutation group acting on $2n$ elements: $P_{2n}$ which is
there to count all the possible ordering of the times $\left\{ \tau_{1},...\tau_{2n}\right\} $.
Now consider the formula in Eq. (\ref{eq:Cumulants}) as a function
of $\delta\in\mathbb{C}$. It is a meromorphic function, and it is
not too hard to see that it has poles of order at most $n$ (this
comes directly from the structure of the denominators). On the other
hand we know that for $\delta$ close to zero the value of $\left\langle \left|Z_{i}\left(\mathrm{T}\right)\right|^{2n}\right\rangle \leq2^{2n}\Lambda^{2n}\mathrm{T}^{2n}$.
This is not obvious from Eq. (\ref{eq:Cumulants}) but is obvious
from the definition of $\left|Z_{i}\left(\mathrm{T}\right)\right|^{2n}$.
As such all the poles in Eq. (\ref{eq:Cumulants}) have to cancel.
Now, schematically a typical term in Eq. (\ref{eq:Cumulants}) may
be written as $\alpha\frac{e^{A\delta\mathrm{T}}}{\delta^{n}}$ (with
$A\in0\cup\mathbb{N}$). As all the poles in $\delta$ must cancel
we may safely replace $\alpha\frac{e^{A\delta\mathrm{T}}}{\delta^{n}}\rightarrow\alpha\frac{\left(A\mathrm{T}\right)^{n}}{n!}$.
From this we see that for large $\mathrm{T}$ to leading order in
$\mathrm{T}$; $\left\langle \left|Z_{i}\left(\mathrm{T}\right)\right|^{2n}\right\rangle \sim\mathrm{T}^{n}$.
The only terms which contribute to order $\mathrm{T}^{n}$ from Eq.
(\ref{eq:Cumulants}) are those $\sim\frac{1}{\delta^{n}}$, or ones
where $\vartheta_{P_{2n}\left(2k\right)}=-\vartheta_{P_{2n}\left(2k-1\right)}$
for $k=1,\,2,...n$. From the fact that the correlation function $e^{-2\Omega_{i}\left|\tau_{1}-\tau_{2}\right|}$
is short ranged and from the fact that the phase factors in Eq. (\ref{eq:Cumulants})
have to cancel pairwise we see that it is good enough to evaluate
$\left\langle \left|Z_{i}\left(\mathrm{T}\right)\right|^{2n}\right\rangle $
in the dipole approximation. From this we see that $\left\langle \left|Z_{i}\left(\mathrm{T}\right)\right|^{2n}\right\rangle \cong n!\left\langle \left|Z_{i}\left(\mathrm{T}\right)\right|^{2}\right\rangle ^{n}$.
These are the moment functions of a complex Gaussian. Repeating the
analysis of Section~\ref{sub:Fluctuating-amplitudes}, we get a power
law decay (for each mode $i$) for the coherence of Majorana qubit,
and Eq. (\ref{eq:telegraph-noise}) follows.

\textit{Eq. (\ref{eq:Cumulants}):} We now wish to derive Eq. (\ref{eq:Cumulants}).
By considering the form of Eq. (\ref{eq:TelegraphNoiseNpoint}) and
the fact that Eq. (\ref{eq:Cumulants}) has a sum over all permutations
of $2n$ elements we see that its enough to derive that:\begin{widetext}\begin{eqnarray}
\int_{0}^{\mathrm{T}}d\tau_{1}e^{\alpha_{1}\tau_{1}}\int_{0}^{\tau_{1}}d\tau_{2}e^{\alpha_{2}\tau_{2}}...\int_{0}^{\tau_{K-1}}d\tau_{K}e^{\alpha_{K}\tau_{K}} & = & \sum_{l=0}^{K}\left\{ \left\{ \left(-1\right)^{K-l}\exp\left(\sum_{j=1}^{l}\alpha_{j}\mathrm{T}\right)\right\} \times\right.\nonumber \\
 & \times & \left.\left(\prod_{j=1}^{l}\frac{1}{\sum_{k=j}^{l}\alpha_{k}}\right)\times\left(\prod_{j=l+1}^{K}\frac{1}{\sum_{k=l+1}^{j}\alpha_{k}}\right)\right\} \label{eq:TimeOrderedIntegral}\end{eqnarray}

To make this formula easier to understand we write it out explicitly
in the case when $K=4$.\begin{eqnarray}
\int_{0}^{\mathrm{T}}d\tau_{1}e^{\alpha_{1}\tau_{1}}\int_{0}^{\tau_{1}}d\tau_{2}e^{\alpha_{2}\tau_{2}}\int_{0}^{\tau_{2}}d\tau_{3}e^{\alpha_{3}\tau_{3}}\int_{0}^{\tau_{3}}d\tau_{4}e^{\alpha_{4}\tau_{4}} & = & \frac{1}{\alpha_{1}\left(\alpha_{1}+\alpha_{2}\right)\left(\alpha_{1}+\alpha_{2}+\alpha_{3}\right)\left(\alpha_{1}+\alpha_{2}+\alpha_{3}+\alpha_{4}\right)}\nonumber \\
 & - & \frac{e^{\alpha_{1}\mathrm{T}}}{\alpha_{1}\alpha_{2}\left(\alpha_{2}+\alpha_{3}\right)\left(\alpha_{2}+\alpha_{3}+\alpha_{4}\right)}\nonumber \\
 & + & \frac{e^{\left(\alpha_{1}+\alpha_{2}\right)\mathrm{T}}}{\left(\alpha_{1}+\alpha_{2}\right)\alpha_{2}\alpha_{3}\left(\alpha_{3}+\alpha_{4}\right)}\label{eq:Example}\\
 & - & \frac{e^{\left(\alpha_{1}+\alpha_{2}+\alpha_{3}\right)\mathrm{T}}}{\left(\alpha_{1}+\alpha_{2}+\alpha_{3}\right)\left(\alpha_{2}+\alpha_{3}\right)\alpha_{3}\alpha_{4}}\nonumber \\
 & + & \frac{e^{\left(\alpha_{1}+\alpha_{2}+\alpha_{3}+\alpha_{4}\right)\mathrm{T}}}{\left(\alpha_{1}+\alpha_{2}+\alpha_{3}+\alpha_{4}\right)\left(\alpha_{2}+\alpha_{3}+\alpha_{4}\right)\left(\alpha_{3}+\alpha_{4}\right)\alpha_{4}}\nonumber \end{eqnarray}

We shall derive Eq. (\ref{eq:TimeOrderedIntegral}) by induction:\begin{eqnarray}
\int_{0}^{\mathrm{T}}d\tau_{1}e^{\alpha_{1}\tau_{1}}...\int_{0}^{\tau_{K-1}}d\tau_{K}e^{\alpha_{K}\tau_{K}} & = & \int_{0}^{\mathrm{T}}d\tau_{1}e^{\alpha_{1}\tau_{1}}\sum_{l=1}^{K}\left\{ -1^{K-l}e^{\sum_{j=2}^{l}\alpha_{j}\tau_{1}}\times\left(\prod_{j=2}^{l}\frac{1}{\sum_{k=j}^{l}\alpha_{k}}\right)\times\left(\prod_{j=l+1}^{K}\frac{1}{\sum_{k=l+1}^{j}\alpha_{k}}\right)\right\} \nonumber \\
 & = & \sum_{l=1}^{K}\left\{ -1^{K-l}\left(e^{\sum_{j=1}^{l}\alpha_{j}\mathrm{T}}-1\right)\left(\prod_{j=1}^{l}\frac{1}{\sum_{k=j}^{l}\alpha_{k}}\right)\times\left(\prod_{j=l+1}^{K}\frac{1}{\sum_{k=l+1}^{j}\alpha_{k}}\right)\right\} \label{eq:InductionStep}\end{eqnarray}

All that remains now is to show that:\begin{equation}
-1^{K}\prod_{i=1}^{K}\frac{1}{\sum_{j=1}^{i}\alpha_{j}}+\sum_{l=1}^{K}\left(-1^{K-l}\left(\prod_{j=1}^{l}\frac{1}{\sum_{k=j}^{l}\alpha_{k}}\right)\times\left(\prod_{j=l+1}^{K}\frac{1}{\sum_{k=l+1}^{j}\alpha_{k}}\right)\right)=0\label{eq:Sum}\end{equation}

\end{widetext}To see this equality consider the left hand side of
Eq. (\ref{eq:Sum}) as a function of $\alpha_{1}\in\mathbb{C}$. This
expression is a meromorphic function $\mathbb{C}\rightarrow\mathbb{C}$
which goes to zero at infinity. By inspection, as a function of $\alpha_{1}$,
it has at most simple poles. It is straightforward to compute the
residues at any of these poles and see that they are all zero, that
is the expression is actually analytic. We can now apply Lioville's
theorem\cite{key-55} to conclude that the function on the left hand
side of Eq. (\ref{eq:Sum}) is identically zero.

\subsubsection{\label{sub:Summation-of-Eq.SumBounds}Summation of Eq. (\ref{eq:SumBounds})
for quadratic Hamiltonians}

We will give an approximate calculation of the sum (\ref{eq:SumBounds})
for tunneling into a 2-D superconductor. To consider a simple example
we will focus on the case where a p-wave superconductor is in close
proximity to a 2-D s-wave superconductor with the chemical potential
of the p-wave superconductor set inside the gap of the s-wave superconductor.
This is a reasonable simplified model for say the surface sates formed
when an STI is placed in proximity to an s-wave superconductor. Furthermore
by taking the limit of a zero gap s-wave superconductor or by ignoring
coherence factors we may model insulators or metals in contact with
p-wave superconductors. We shall assume a constant point tunneling
contact so that the relevant tunneling Hamiltonian may be written
as:\begin{equation}
\int d^{2}r\mathbb{T}\left(\Psi_{\mathrm{pw}}^{\dagger}\left(r\right)\Psi_{\mathrm{sw}\uparrow}\left(r\right)+\Psi_{\mathrm{sw}\uparrow}^{\dagger}\left(r\right)\Psi_{\mathrm{pw}}\left(r\right)\right)\label{eq:BasicTunnelingHamiltonian}\end{equation}
 This form comes from the fact that for a p-wave superconductor the
vortex is in one spin species only, say spin up.

We begin with a review of the relevant wavefunctions for zero modes
of a p-wave superconductor. The eigenvalues of our Hamiltonian correspond
to solutions of the following BdG equation:\begin{widetext}

\begin{equation}
\mathrm{\left(\begin{array}{cc}
-\frac{\nabla^{2}}{2m}-\mu & \frac{1}{2}\left\{ \Delta\left(\vec{r}\right),p_{x}-ip_{y}\right\} \\
\frac{1}{2}\left\{ \Delta^{*}\left(\vec{r}\right),p_{x}+ip_{y}\right\}  & \frac{\nabla^{2}}{2m}+\mu\end{array}\right)\left(\begin{array}{c}
u\\
v\end{array}\right)=\epsilon\left(\begin{array}{c}
u\\
v\end{array}\right)}\label{eq:BdGPWAVE}\end{equation}
 Here $\Delta\left(\vec{r}\right)=\exp\left(i\theta\right)\Delta\left(\left|\vec{r}\right|\right)$,
with $\Delta\left(\left|\vec{r}\right|\right)=\frac{\left|\vec{r}\right|}{\xi}\Delta_{\infty}$
for $\left|\vec{r}\right|\leq\xi$ and $\Delta\left(\left|\vec{r}\right|\right)=\Delta_{\infty}$
for $\left|\vec{r}\right|\geq\xi$ (we have neglected an irrelevant
overall phase factor). Here $\xi$ is the penetration depth and $\Delta_{\infty}$
is the magnitude of the order parameter far from the vortex. From
previous studies \cite{key-40,key-41}, for rotationally symmetric
type II superconducting vortices, we know that there is a zero mode
for the Hamiltonian given in Eq. (\ref{eq:BdGPWAVE}). It is given
by $\gamma=\int d^{2}r\left(u_{0}\left(r\right)\Psi\left(\overrightarrow{r}\right)+v_{0}(r)\Psi^{\dagger}\left(r\right)\right)$
with:\begin{equation}
\begin{array}{l}
\left(\begin{array}{c}
u_{0}\left(r\right)\\
v_{0}\left(r\right)\end{array}\right)\cong\frac{N}{\sqrt{2}}J_{0}\left(k_{F}r\right)\exp\left(-\chi\left(r\right)\right)\left(\begin{array}{c}
1+i\\
1-i\end{array}\right)\end{array}\label{eq:Wavefunctions}\end{equation}

\end{widetext} Here $k_{F}=\sqrt{2m\mu}$ is the Fermi wavevector,
$J_{0}\left(k_{F}r\right)$ is the l'th Bessel function and $\chi\left(r\right)=\frac{m}{k_{F}}\int_{0}^{r}\Delta\left(r\right)$.
Where $\Delta\left(r\right)$ is the position dependent order parameter.
Furthermore a good approximate value for the normalization constant
is given by $N\cong0.06\left(\frac{k_{F}}{\xi}\right)$ (see \cite{key-40}).

Next we will recall the form of the wavefunctions for an s-wave superconductor.
For s-wave superconductors we may write Bogolubov de Gennes equations
in the form:\begin{equation}
\left(\begin{array}{cc}
-\frac{\nabla^{2}}{2m}-\widetilde{\mu} & \widetilde{\Delta}\\
\widetilde{\Delta}^{*} & \frac{\nabla^{2}}{2m}+\widetilde{\mu}\end{array}\right)\left(\begin{array}{c}
f\left(r\right)\\
g\left(r\right)\end{array}\right)=E\left(\begin{array}{c}
f\left(r\right)\\
g\left(r\right)\end{array}\right)\label{eq:BdG}\end{equation}

Here the top component represents creation operators for spin up while
the bottom component represents annihilation operators for spin down
fermions; $\widetilde{\mu}$ and $\widetilde{\Delta}$ are the chemical
potential and the gap of the s-wave superconductor. Furthermore a
similar equation may be written with the spins interchanged and $\widetilde{\Delta}\rightarrow-\widetilde{\Delta}$.
We will place the origin of co-ordinates at the center of the vortex
in the p-wave superconductor. Solutions for this equation are of the
form:\begin{equation}
\mathrm{\left(\begin{array}{c}
f^{\left(+,-\right)}\left(r\right)\\
g^{\left(+,-\right)}\left(r\right)\end{array}\right)=\frac{1}{\mathcal{C}}\left(\begin{array}{c}
A^{\left(+,-\right)}e^{il\theta}J_{l}\left(qr\right)\\
B^{\left(+,-\right)}e^{il\theta}J_{l}\left(qr\right)\end{array}\right)}\label{eq:Eigenfunction}\end{equation}

Here $\mathcal{C}$ is a size dependent normalization constant with
$\frac{1}{\mathcal{C}}\cong\frac{\pi q}{R}$ (where $R$ is the system
radius). Eigenenergies and eigenfunctions are now given by: \begin{equation}
\left\{ \mathrm{\begin{array}{l}
E^{\left(+,-\right)}=\pm\sqrt{\left(\frac{q^{2}}{2m}-\widetilde{\mu}\right)^{2}+\widetilde{\Delta}^{2}}\\
\left(A^{+},\, B^{+}\right)=\left(\cos\left(\theta/2\right)\exp\left(i\widetilde{\varphi}\right),\:\sin\left(\theta/2\right)\right)\\
\left(A^{-},\, B^{-}\right)=\left(-\sin\left(\theta/2\right)\exp\left(i\widetilde{\varphi}\right),\:\cos\left(\theta/2\right)\right)\end{array}}\right.\label{eq:EigenvalueSolutions}\end{equation}
 Here $\mathrm{\tan\left(\theta\right)=\frac{\frac{q^{2}}{2m}-\widetilde{\mu}}{\widetilde{\Delta}}}$,
$\mathrm{\frac{\widetilde{\Delta}}{\widetilde{\Delta}^{*}}=\exp\left(i2\widetilde{\varphi}\right)}$
and $J_{l}$ are the l'th Bessel functions. There are completely analogous
equations for the opposite spin, with appropriate sign and phase changes.
Using Eq. (\ref{eq:Tunneling}) as well as the symmetry between the
upper and lower component of the solution for the zero mode, see Eq.
(\ref{eq:Wavefunctions}) and various symmetries between the spin
species we see that various trig functions (such as the sine, cosine
and exponential appearing in the solution of Eq. (\ref{eq:Eigenfunction})
above) cancel out. By taking the thermodynamic limit we can convert
the sum (\ref{eq:SumBounds}) into an integral of the form: $\sum_{i=1}^{N}\frac{\left|\Gamma_{i}\right|^{2}}{\epsilon_{i}^{2}}\cong$
\begin{widetext}\begin{equation}
8\pi\int_{0}^{\infty}dq\left(\frac{1}{\left(\left(\widetilde{\mu}-\mu\right)+\sqrt{\left(\frac{q^{2}}{2m}-\widetilde{\mu}\right)^{2}+\widetilde{\Delta}^{2}}\right)^{2}}+\frac{1}{\left(\mathrm{\left(\widetilde{\mu}-\mu\right)-\sqrt{\left(\frac{q^{2}}{2m}-\widetilde{\mu}\right)^{2}+\widetilde{\Delta}^{2}}}\right)^{2}}\right)N^{2}\left|\mathbb{T}\int_{0}^{\infty}drru_{0}\left(r\right)J_{0}\left(qr\right)\right|^{2}\label{eq:Integral}\end{equation}

We note that because of rotational invariance only $J_{l=0}$ terms
contribute to the sum. Here $u_{0}$ is the upper component of the
Majorana mode wavefunction (Eq. (\ref{eq:Wavefunctions})). We wish
to evaluate the integral given in Eq. (\ref{eq:Integral}) above.
We will begin by evaluating $\int_{0}^{\infty}drru_{0}\left(r\right)J_{0}\left(qr\right)$.
As a first step we will use the approximate relation that: $u_{0}\left(r\right)\cong N\exp\left(-\frac{\Delta}{k_{F}\xi}r^{2}\right)J_{0}\left(k_{F}r\right)$
(see Eq. (\ref{eq:Wavefunctions}) and discussion that immediately
follows). Next we write that:\begin{eqnarray}
\int_{0}^{\infty}drru_{0}\left(r\right)J_{0}\left(qr\right) & = & \frac{N}{2\pi}\int_{-\infty}^{\infty}dx\int_{-\infty}^{\infty}dy\exp\left(-\frac{\Delta}{k_{F}\xi}r^{2}\right)J_{0}\left(k_{F}r\right)J_{0}\left(qr\right)\nonumber \\
 & = & \frac{N}{\left(2\pi\right)^{3}}\int_{-\infty}^{\infty}dx\int_{-\infty}^{\infty}dy\exp\left(-\frac{\Delta}{k_{F}\xi}r^{2}\right)\int_{0}^{2\pi}d\theta_{1}e^{-i\overrightarrow{k_{F}}\left(\theta_{1}\right)\cdot\vec{r}}\int_{0}^{2\pi}d\theta_{2}e^{-i\vec{q}\left(\theta_{2}\right)\cdot\vec{r}}\label{eq:BesselIntegral}\\
 & = & \frac{N}{\left(2\pi\right)^{3}}\int_{0}^{2\pi}d\theta_{1}\int_{0}^{2\pi}d\theta_{2}\exp\left(-\frac{k_{F}\xi}{4\Delta}\left(\overrightarrow{k_{F}}\left(\theta_{1}\right)+\vec{q}\left(\theta_{2}\right)\right)^{2}\right)\nonumber \\
 & = & \frac{N}{2\pi^{2}}\int_{-1}^{1}\frac{dx}{\sqrt{1-x^{2}}}\exp\left(-\frac{k_{F}\xi}{4\Delta}\left(k_{F}^{2}+q^{2}+2qk_{F}x\right)\right)\nonumber \\
 & = & \frac{N}{2\pi}\times I_{0}\left(\frac{q\xi}{2\Delta}\right)\exp\left(-\frac{k_{F}\xi}{4\Delta}\cdot\left(k_{F}^{2}+q^{2}\right)\right)\nonumber \\
 & \cong & \frac{N}{2\pi}\times\sqrt{\frac{\Delta}{\pi q\xi}}\times\exp\left(-\frac{k_{F}\xi}{4\Delta}\left(q-k_{F}\right)^{2}\right)\nonumber \end{eqnarray}
 \end{widetext} Here $\overrightarrow{k_{F}}\mathrm{\left(\theta_{1}\right)}$
is a vector with magnitude $k_{F}$ and direction $\theta_{1}$ along
the x-axis and similarly for $\vec{q}\left(\theta_{2}\right)$. In
the second line we have used a representation of the bessel function:
$J_{0}\left(qr\right)=\frac{1}{2\pi}\int_{0}^{2\pi}d\theta e^{-i\vec{q}\left(\theta\right)\cdot\vec{r}}$
and $\vec{r}$ is along the y-axis. Here $I_{0}$ is a modified Bessel
function of zeroth order and in the last step we have used an asymptotic
form of the modified Bessel function $I_{0}\left(\frac{q\xi}{2\Delta}\right)\cong\sqrt{\frac{\Delta}{\pi q\xi}}\exp\left(-\left(\frac{q\xi}{2\Delta}\right)^{2}\right)$.
This asymptotic form fails near $q=0$ where it should be replaced
by $I_{0}\left(\frac{q\xi}{2\Delta}\right)\cong1+\frac{1}{4}\left(\frac{q\xi}{2\Delta}\right)^{2}+..$.
It is straight forward to check that this correction does not effect
the final answer see Eq. (\ref{eq:LimitingValues}) below. Indeed
because of the exponential decay we may safely approximate: \begin{equation}
\int_{0}^{\infty}drru_{0}\left(r\right)J_{0}\left(qr\right)\cong\left\{ \begin{array}{cc}
\frac{N}{2\pi}\sqrt{\frac{\Delta}{\pi q\xi}} & \,\left(q-k_{F}\right)\leq\frac{\Delta}{k_{F}\xi}\\
0 & \,\left(q-k_{F}\right)\geq\frac{\Delta}{k_{F}\xi}\end{array}\right.\label{eq:LimitingValues}\end{equation}

From this we see that the integral given in Eq. (\ref{eq:Integral})
above has effectively a finite range of definition and no singularities.
As such it is clearly finite. Very similar arguments may be used to
show that the sum (\ref{eq:SumBounds}) is bounded for tunneling contact
with any gaped material such as an insulator with the chemical potential
of the p-wave superconductor lying within the gap. Indeed quite generically
for an itinerant system we may write the Hamiltonian as $H=-\mathrm{\frac{\nabla^{2}}{2m^{*}}}+...$
which means that the eigenvectors of $H$ are similar to those of
an s-wave superconductor so the integrand in Eq. (\ref{eq:Integral})
above also has exponential decay for large momentum as the solutions
of $H\left|\Psi\right\rangle =E\left|\Psi\right\rangle $ would behave
almost like Bessel functions. Because of the gap condition there will
be no finite momentum divergences either, leading to a finite integral.
This argument may be extended to models with band structure. By {}``folding
out'' appropriate bands from the first brillouin we may convert the
sum $\sum_{\delta}\int\int_{BZ}\left(\frac{\Gamma_{k\delta}}{\epsilon_{k\delta}}\right)^{2}$
(where the integral is over the first Brillouin zone) into an integral
over all of k-space $\rightarrow\int\int d^{2}k\left(\frac{\Gamma_{k\delta}}{\epsilon_{k\delta}}\right)^{2}$.
As any possible divergence would come from high energy bands where
the dispersion is essentially quadratic and the wavefunction is essentially
of the continuum model, we may reduce the problem to a previously
solved case.


\begin{thebibliography}{50}
\bibitem{key-43} More generally we will ignore any spin structure
of the Majorana modes from now on. Such a structure merely leads to
a various matrix elements between the Majorana mode and surrounding
states. These matrix elements may be absorbed into the definition
of $\Gamma_{i}$ (see Eq. (\ref{eq:HamiltonianQuadraticTunneling})).
\bibitem{key-89} Indeed all possible operators acting on the subspace
spanned by the zero modes of the Hamiltonian may be written as sums
and products of Majorana modes.
\bibitem{key-45} Some readout schemes do not make the distinction
between states inside and outside the coding space.
\bibitem{key-51} Though we do not use this result we note that a
symmetry argument shows that $\left\langle \prod_{i=1}^{N}\Gamma\left(\tau_{i}\right)\right\rangle =0$
for odd $N$.
\end{thebibliography}
\end{document}